\documentclass[12pt,sort&compress]{elsarticle}

\usepackage{afterpage}
\usepackage{enumitem}
\usepackage{mystyle}
\usepackage{algpseudocode}
\usepackage{algorithm}
\usepackage{amsmath}
\usepackage{amssymb}
\usepackage{subfiles}
\graphicspath{{Images/}{../Images/}}
\usepackage{subfigmat} 
\usepackage{relsize}
\usepackage{pst-all}
\usepackage[normalem]{ulem}
\usepackage{cancel}
\usepackage{caption}
\usepackage{cleveref}
\usepackage{comment}

\newcommand\figref[1]{Figure \ref{fig:#1}} 
\newcommand\tabref[1]{Table \ref{tab:#1}} 
\newcommand\eref[1]{Eq. (\ref{eq:#1})} 
\newcommand\p[1]{\partial{#1}}

\begin{document}

\begin{frontmatter}


\title{Nonintrusive projection-based reduced order modeling using stable learned differential operators}

\author[LANL]{Aviral Prakash\corref{cor1}}
\ead{aviralp@lanl.gov}
\author[CMU]{Yongjie Jessica Zhang}
\cortext[cor1]{Corresponding author. Most of the work was done when Dr. Prakash was at Carnegie Mellon University.}

\address[LANL]{Theoretical Division, Los Alamos National Laboratory, Los Alamos, NM 87545, USA}
\address[CMU]{Department of Mechanical Engineering, Carnegie Mellon University, Pittsburgh, PA 15213, USA}

\begin{abstract}
    Nonintrusive projection-based reduced order models (ROMs) are essential for dynamics prediction in multi-query applications where access to the source of the underlying full order model (FOM) is unavailable; that is, FOM is a black-box. This article proposes a \textit{learn-then-project} approach for nonintrusive model reduction. In the first step of this approach, high-dimensional stable sparse learned differential operators (S-LDOs) are determined using the generated data. In the second step, the ordinary differential equations, comprising these S-LDOs, are used with suitable dimensionality reduction and low-dimensional subspace projection methods to provide equations for the evolution of reduced states. This approach allows easy integration into the existing intrusive ROM framework to enable nonintrusive model reduction while allowing the use of Petrov-Galerkin projections. The applicability of the proposed approach is demonstrated for Galerkin and LSPG projection-based ROMs through three numerical experiments: 1-D advection equation, 1-D Burgers equation and 2-D advection equation. The results indicate that the proposed nonintrusive ROM strategy provides accurate and stable dynamics prediction.
\end{abstract}


\end{frontmatter}

\section{Introduction}

Advancements in numerical algorithms and computational hardware have enabled large-scale accurate simulations for multi-physics applications. However, these high-fidelity solutions of partial differential equations (PDEs), referred to as full order models (FOMs), are typically associated with high computational costs that make them infeasible from long-term dynamics prediction, real-time estimation and multi-query applications such as design space exploration, design optimization, uncertainty quantification and parameter estimation. Over the past two decades, there has been a growing interest in reduced order models (ROMs) for such scenarios, as these models have the potential to provide accurate physics predictions at a much lower cost than FOMs. 

ROMs achieve relatively low computational costs using high-fidelity generated data and employing an offline-online stage decomposition. This decomposition ensures that expensive computational operations are restricted to the offline stage, which is carried out only once. In contrast, computationally inexpensive operations are performed in the online stage and can be carried out for multiple instances without incurring massive costs. In the first step of the offline stage, the data generated from experiments or FOMs is used to identify an optimal solution representation that reduces the dimensionality of the system to provide reduced states. Several approaches proposed in the past used linear or affine decomposition to determine the optimal solution representation \cite{Lumley1967, Hesthaven2016}. Recently, there has been a growing interest in nonlinear solution representations \cite{Lee2020, Kim2022, Puri2024}, especially for applications exhibiting a slow decay of Kolmogorov $n$-width. This article will restrict discussion to linear solution representation using proper orthogonal decomposition (POD) \cite{Lumley1967, Aubry1988} as it is the most commonly used dimensionality reduction method. The offline stage often involves a second step where relevant evolution equations of reduced states are derived. This step becomes a part of the online stage of ROMs for PDEs with non-polynomial nonlinearity or when nonlinear dimensionality reduction methods are used. A common strategy for deriving these equations involves projecting PDEs, or the resulting ordinary differential equations (ODEs), onto a low-dimensional subspace or solution manifold. ROMs involving this projection step are often known as projection-based ROMs. There are several strategies for projections, such as Galerkin projection \citep{Aubry1988} and Petrov-Galerkin projection \cite{Carlberg2011, Carlberg2017, Parish2020, Parish2023}, that have been proposed over the years. In the online stage, the derived equations of reduced states following the projection step are marched in time using appropriate time integration schemes. As the dimensionality of these equations is significantly lower than FOMs, the online stage of ROMs is computationally inexpensive and can be affordably realized multiple times in a multi-query application.
    
While projection-based ROMs have been demonstrated to be an efficient simulation strategy for several PDEs, these ROMs require an intimate knowledge of the physical equations and discretization methods. As a result, these ROMs are commonly referred to as intrusive ROMs. Intrusive ROMs are undesirable for several scenarios. For example, developing reduced-order models based on commercial simulation software or for national security applications will be challenging as access to the underlying code is not available. Furthermore, as intrusive ROMs are closely tied to underlying FOM, benchmarking different ROM strategies and validation between different research groups becomes arduous. Lastly, intrusive ROMs have greater implementation complexity as intimate knowledge of the FOM code is required and may hinder the widespread application of novel ROM strategies. Due to these limitations, there has been growing interest in nonintrusive ROMs that do not require development within the FOM code \cite{Peherstorfer2016, Hesthaven2018, Fries2022, Padovan2024}. Some nonintrusive ROMs are purely data-driven; that is, they use generated data to directly determine equations of reduced states without enforcing any knowledge of governing physical equations \cite{Rowley2009, Schmid2010, Schmid2022, Hesthaven2018, Champion2019, Fries2022}. Despite the nonintrusive nature of such ROMs, a lack of dependence on the governing physical equations diminishes the confidence of these models for scenarios not used in the learning process, such as extrapolation in time or parameter space. Furthermore, like in discretization methods, consistency is also an essential property for ROMs, which, along with a stable ROM formulation, may ensure convergence of predictions to projected FOM results \cite{Kalashnikova2010}. Despite the importance of this attribute and widespread attention in discretization literature, there is only a limited discussion in the context of ROMs \cite{Kalashnikova2010, Ingimarson2022, Parish2023}. Nonintrusive ROMs that determine underlying equations of reduced states using data may not be consistent with FOMs, thereby lacking convergence of ROM results to those from FOMs. Conversely, standard projection-based ROMs are more consistent with the underlying FOM while exhibiting the undesirable property of intrusiveness. 

In this article, we consider nonintrusive projection-based ROMs for scenarios where the relevant governing physical equations are available while the knowledge of underlying numerical methods and access to the software framework to solve these equations is unavailable. ROMs applicable in such scenarios are often called glass-box ROMs \cite{McQuarrie2021}. As the equations of reduced states are derived using governing physical equations for projection-based ROMs, they are more consistent than purely data-driven nonintrusive ROMs. Operator inference is a commonly used nonintrusive model reduction method that has been demonstrated to be robust for several applications \cite{Peherstorfer2016, Kramer2024}. This method follows a \textit{project-then-learn} approach for nonintrusive model reduction. In this approach, the first step requires knowledge of physical equations and Galerkin projection to identify the form of the equation of reduced states. In the second step, ROM operators are determined by solving a regression problem. Over the past years, operator inference has been refined to improve accuracy and stability by accounting for Markovian effects \cite{Peherstorfer2020}, adding regularization \cite{McQuarrie2021} and preserving physical properties \cite{Sawant2023, Sharma2022}. A recent review of operator inference \cite{Kramer2024} provided a detailed discussion on recent advances in this area. Despite the positive attributes, operator inference acts as a standalone ROM approach that cannot be integrated with existing ROMs. To the authors' knowledge, there is no direct extension of the operator inference approach for Petrov-Galerkin projection-based ROMs, which have been demonstrated superior to Galerkin-projection for complex nonlinear PDE problems \cite{Carlberg2017, Grimberg2020, Parish2023}. 

The goal of this article is to propose an alternate nonintrusive projection-based model reduction approach that overcomes the abovementioned shortcomings. The proposed nonintrusive model reduction follows \textit{learn-then-project} approach. The first step in this approach involves learning high-dimensional differential operators from data. The second step involves applying Galerkin or Petrov-Galerkin projections to obtain equations of reduced states. Learning high-dimensional operators is a challenging problem that can be computationally expensive. Furthermore, desirable properties of these operators, such as stability, are commonly not enforced \cite{Schumann2022, Schumann2023, Gkimisis2023, Gkimisis2024}. To overcome these issues, we recently proposed an approach for obtaining stable sparse learned differential operators (S-LDOs) from data in \cite{Prakash2024b}. This strategy uses the Gershgorin circle theorem to identify local constraints while ensuring the positive definiteness of the global operator. These local constraints are used with local regression problems to identify local operators that yield stable global differential operators. As the local operators are selected to be sparse, this approach allows determining differential operators with local support, thereby reducing the cost of the regression problem. The discussion in \cite{Prakash2024b} was limited to the mathematical formulation of S-LDOs, while its applicability for nonintrusive model reduction was not shown. This article demonstrates that S-LDOs enable effective adoption of the \textit{learn-then-project} approach for nonintrusive projection-based ROMs. Nonintrusive ROMs using S-LDOs are demonstrated to be suitable even for Petrov-Galerkin projection-based ROMs, which may not be possible with operator inference. Furthermore, we prove that using S-LDOs with Galerkin projection yields theoretically stable ROMs following Lyapunov's indirect method. The applicability of S-LDOs in nonintrusive projection-based ROMs is demonstrated through three numerical examples: 1-D scalar advection equation, 1-D Burger equation and 2-D advection equation. 

The detailed outline of this article is given below. In Section \ref{sec:MathFrame}, we first provide the relevant mathematical background on differential operators and desirable stability properties. Second, we discuss the formulation of S-LDOs for linear and nonlinear problems. In Section \ref{sec:PROM}, we introduce the construction of projection-based ROMs focusing on Galerkin and least-squares Petroc Galerkin (LSPG) projections. We discuss the incorporation of S-LDOs to enable nonintrusiveness in projection-based ROMs and prove the stability of Galerkin projection-based ROMs when S-LDOs are used as differential operators. We provide the implementation details of the proposed nonintrusive ROM approach. We also highlight the main differences between nonintrusive ROMs obtained using operator inference and S-LDOs. In Section \ref{sec:Results}, we demonstrate the applicability of S-LDO enabled nonintrusive Galerkin and LSPG projection-based ROMs for linear and nonlinear problems. In Section \ref{sec:Conclusions}, we gather the major conclusions and underline future research directions enabled by the current work.

\section{Mathematical background}
\label{sec:MathFrame}

In this article, we consider PDEs of the form
\begin{equation}
    \frac{\p u}{\p t} + \mathcal{L} u + \mathcal{N} (u) = 0,
\end{equation}
where $u (x, t)$ is the solution, $\mathcal{L}$ is linear differential operator and $\mathcal{N}$ is the nonlinear differential operator. This equation is subject to an initial condition $u(x, 0) = u_0$ and boundary conditions on the boundary of the problem domain $\Gamma \subset \Omega$. While the mathematical formulation is explained in a 1-D setting, the proposed methods directly apply to higher-dimensional problems. In Section \ref{sec:DiscreteOpt}, we discuss important properties that must be satisfied by differential operators in the discrete form of PDEs. In Section \ref{sec:SLDO}, an approach proposed in \cite{Prakash2024b} for obtaining stable sparse differential operators is discussed.

\subsection{Discrete differential operators and stability}
\label{sec:DiscreteOpt}

Using a suitable spatial discretization method, the PDE can be written in discrete form represented by a set of ODEs following the method of lines \cite{LeVeque2007}. This discrete form is represented as
\begin{equation}
    \frac{d \bm{u}}{d t} + \bm{L} \bm{u} + \bm{N} (\bm{u}) = 0,
    \label{eq:PDE_def}
\end{equation}
where $\bm{L} \in \mathbb{R}^{n\times n}$ and $\bm{N}:  \mathbb{R}^n \to \mathbb{R}^n$ are linear and nonlinear differential operators, while $\bm{u} \in \mathbb{R}^n$ is a vector comprising of $n$ discrete degrees of freedom which depend on the selected discretization scheme. While the accuracy of the spatial discretization schemes is the biggest consideration, these schemes are commonly chosen to ensure that these differential operators are sparse, also referred to as local support, for enabling faster solutions using optimized sparse linear algebra algorithms and libraries \cite{Duff2002}. 

In addition to accuracy and sparsity, stability of the discretized PDEs, or the set of resulting coupled ODEs, is also an important consideration while selecting an appropriate discretization scheme. The stability of discretization schemes is typically addressed using Von-Neumann analysis for finite difference schemes \cite{Charney1950, Isaacson1994} and using the notion of energy stability in finite element context \cite{Evans2023}. For example, as demonstrated through Von Neumann's analysis, unwinding is common in finite difference \cite{LeVeque2007} and finite volume schemes \cite{LeVeque2002} to ensure stability for hyperbolic problems. Similarly, stabilized methods for finite elements are common for simulating hyperbolic problems and are often developed in conjunction with the notion of energy stability \cite{Evans2023}. While these notions of stability are adequate for assessing discretization schemes, this article instead focuses on discrete differential operators, which are finite-dimensional. Therefore, we instead consider the theory of stability borrowed from dynamical systems literature in conjunction with the semi-discrete form of \eref{PDE_def}. Considering a linear system of ODEs of the form
\begin{equation}
    \frac{d \bm{u}}{d t} = \bm{A} \bm{u},
    \label{eq:PDE_def_linear}
\end{equation}
the stability of this system follows

\bigskip
\noindent \textbf{Definition 2.1}: The linear system in \eref{PDE_def_linear} is referred to as stable if the real part of all eigenvalues of $\bm{A}$ is non-positive. The system is commonly called asymptotically stable if the real part of all eigenvalues of $\bm{A}$ is negative. The system is unstable if $\bm{A}$ has an eigenvalue with a positive real part. 
\bigskip

Therefore, following Definition 2.1, \eref{PDE_def} without the nonlinear term, is stable if $\bm{L}$ is positive semi-definite. For nonlinear systems, in this article, we consider the definition of stability based on linearized equations
\begin{equation}
    \frac{d \bm{u}}{d t} + \bm{N}^L \bm{u} = 0,
    \label{eq:PDE_def_taylor}
\end{equation}
\noindent where
\begin{equation}
    \bm{N}^L = \Big\vert \frac{\partial (\bm{L} \bm{u} + \bm{N} (\bm{u}))}{\partial \bm{u}} \Big\vert_{\bm{u_0}}
    \label{eq:NonlinConstr}
\end{equation}
is the linearized operator and $\vert \cdot \vert_{\bm{u_0}}$ implies the linearized operator is evaluated at an equilibrium point $\bm{u}_0$. Following Lyapunov's indirect method \cite{Sastry1999}, the system of equations in \eref{PDE_def} is stable if the linearized operator $\bm{N}^L$ is positive definite. Note that the stability of nonlinear systems is not ensured if any eigenvalues of $\bm{N}^L$ are zero due to the influence of neglected terms during linearization on the solution behavior. There are other notions of nonlinear stability following Lyapunov's direct method \cite{Rouche1977}, however these are not considered in this article. 

\subsection{Stable learned differential operators (S-LDOs)}
\label{sec:SLDO}

Decades of research in discretization schemes have matured these numerical methods to enable fast and accurate simulation of multiphysics problems. However, in nonintrusive reduced order modeling, these discretization schemes are considered unknown as access to them is unavailable. This situation often arises when working with commercial or national security-related simulation software. We can rely on simulation data to identify the underlying in such settings. 

An accurate unknown discretization or discrete differential operators can be identified from simulation data by posing it as a regression problem \cite{Gkimisis2024, Schumann2022, Schumann2023}. In such scenario, instead of \eref{PDE_def}, we consider
\begin{equation}
    \frac{d \bm{u}}{d t} + \bm{L}^m \bm{u} + \bm{N}^m (\bm{u}) = 0,
    \label{eq:PDE_def_discrete}
\end{equation}
where $\bm{L}^m$ and $\bm{N}^m$ are unknown or modeled linear and nonlinear differential operators determined from high-fidelity simulation data. Instead of learning a high-dimensional global dynamical system, these methods learn sparse differential operators by subdividing the differential operator learning problem to the problem of learning individual ODEs. The ODE for the $i^{th}$ degree of freedom in \eref{PDE_def_discrete} is 
\begin{equation}
    \frac{d u_i}{d t} + (\bm{L}^{m,i})^T \bm{u}_{\Omega^l_i} + \bm{\hat{N}}^{m,i} (\bm{u}_{\Omega^n_i}) = 0,
    \label{eq:PDEeq3}
\end{equation}
where $\bm{L}^{m,i} \in \mathbb{R}^{s_l}$ is the local linear operator which contributes to the $i^{th}$ row of $\bm{L}^m$, superscript $T$ indicates the transpose, $s_l$ is the size of the local linear stencil, $\bm{\hat{N}}^{m,i}: \mathbb{R}^{s_n} \to \mathbb{R}$ is the local nonlinear operator and $s_n$ is the size of the local nonlinear stencil. To simplify the explanation, we transform the nonlinear term in \eref{PDEeq3} to a matrix-vector product, which is similar to the linear term. The resulting equation is
\begin{equation}
   \frac{d u_i}{d t} + (\bm{L}^{m,i})^T \bm{u}_{\Omega^l_i} + (\bm{N}^{m,i})^T \bm{z} (\bm{u}_{\Omega^n_i}) = 0,
    \label{eq:PDEeq4}
\end{equation}
where $\bm{N}^{m,i} \in \mathbb{R}^{s_n}$ is the nonlinear differential operator in a vector form, $s_n$ is the nonlinear stencil size and $\bm{z} (\bm{u}_{\Omega^n_i}) \in \mathbb{R}^{s_n}$ represents nonlinear products of $\bm{u}_{\Omega^n_i}$. The local differential operators in \eref{PDEeq4} can be obtained from data by solving a regression problem proposed in \cite{Gkimisis2024, Schumann2022, Schumann2023}. While these differential operators are shown to yield accurate results for dynamics prediction, they are not designed to be stable. For example, numerical instabilities were observed for 1-D hyperbolic problems \cite{Prakash2024b}. Therefore, these operators do not guarantee accurate and stable results for dynamics forecasting problems. While Tikhonov regularization has been shown to improve stability \cite{McQuarrie2021, Sawant2023}, this regularization may not guarantee stable operators as shown in \cite{Prakash2024b}. There has been recent work on learning stable dynamical systems using constrained regression \cite{Boots2007} and matrix parameterization \cite{Goyal2023, Goyal2024}. While these methods suit low-dimensional systems, the computational cost is extremely high for high-dimensional systems, such as \eref{PDE_def} for common 2-D/3-D PDE problems. This behavior is attributed to the $O(n^2)$ cost of the unconstrained regression problem. In the presence of constraints, this cost will be even higher. Furthermore, the learned differential operators are not sparse, thereby losing out on local support property and the associated efficiency of such sparse systems. 

To overcome this drawback, a novel strategy that learns stable sparse differential operators, commonly referred to as S-LDOs, was proposed in \cite{Prakash2024b}. This strategy poses the identification of differential operators as a constrained regression problem defined as follows:

\bigskip
\noindent \textit{Given high-fidelity data, $\bm{u} \in \mathbb{R}^n$ and $\dot{\bm{u}} = \frac{d \bm{u}}{d t} \in \mathbb{R}^n$ at time instances $t = t_j$ for $j = 1, \cdot \cdot \cdot,\; n_t$ and solution stencils for the two operators $\bm{u}_{\Omega^l_i} \in \mathbb{R}^{s_l}$ and $\bm{u}_{\Omega^n_i} \in \mathbb{R}^{s_n}$, find differential operators $\bm{L}^{m,i}$ and $\bm{N}^{m,i}$, subject to the objective function}
\begin{equation}
    \underset{\tilde{\bm{L}}^i, \; \tilde{\bm{N}}^i}{\text{min}} \; \Big\vert \Big\vert \dot{u}_i (t) + (\tilde{\bm{L}}^i)^T \bm{u}_{\Omega^l_i} (t) + (\tilde{\bm{N}}^i)^T \bm{z} (\bm{u}_{\Omega^n_i} (t)) \Big\vert \Big\vert_2^2 \quad \forall \; i = 1, \cdot \cdot \cdot, n
    \label{eq:RegProbConstrainedFinalNonLin}
\end{equation}
\textit{satisfying the constraint}
\begin{equation}
    N^{m,L,i}_{i} - \sum_{j \neq i} \vert N^{m,L,i}_{j} \vert > 0 \quad \forall \; i = 1, \cdot \cdot \cdot, n,
    \label{eq:ConstrainedFinalNonLin}
\end{equation}
\textit{which is the localized form of \eref{NonlinConstr} where $N^{m,L,i}_{j}$ is the $j^{th}$ element of $\bm{N}^{m,L,i}$. The local terms $\bm{N}^{m,L,i}$ are assembled to form $\bm{N}^{m,L}$}, which is the modeled counterpart of $\bm{N}^L$. 

\bigskip
\noindent Following this regression problem, the resulting modeled differential operators $\bm{L}^m$ and $\bm{N}^m$ can be obtained by assembling local differential operators $\bm{L}^{m,i}$ and $\bm{N}^{m,i}$. Another strategy involves not assembling these operators and instead handling them in a matrix-free format, which can be efficient for heterogeneous computer architectures with hardware accelerators. The resulting differential operators $\bm{L}^m$ and $\bm{N}^m$ are sparse and locally stable, therefore called stable learned differential operators or S-LDOs. Note that the sparsity in S-LDOs is fixed using a user-defined stencil size parameter and does not use $L_1$ regularization techniques to identify a sparse stencil. The next section will combine these S-LDOs with projection-based reduced order modeling methods to enable nonintrusive model reduction.

\section{\textit{Learn-then-project} approach for nonintrusive model reduction}
\label{sec:PROM}

In the previous section, we discussed the development of S-LDOs which is the key ingredient for \textit{learn-then-project} approach to nonintrusive model reduction. This section introduces basic mathematical formulation for projection-based reduced order modeling while demonstrating the utility of S-LDOs in enabling nonintrusiveness. Section \ref{sec:POD} describes the mathematical formulation of POD for linear dimensionality reduction for ROMs. Section \ref{sec:DiscreteProj} introduces Galerkin and Petrov-Galerkin projections to obtain equations of reduced states. Section \ref{sec:SLDO-ROM} formalizes using S-LDO for nonintrusive model reduction while providing mathematical proofs of stability for ROMs with S-LDOs and discrete Galerkin projection. Section \ref{sec:Algo} highlights the numerical algorithm for implementing the proposed approach for nonintrusive model reduction. Lastly, Section \ref{sec:OpInfComp} highlights the key differences between the proposed nonintrusive model reduction using S-LDOs and Operator Inference, a commonly used nonintrusive model reduction method. In this article, the solution is represented as a finite-dimensional vector, which enables discrete projection to determine equations of reduced states. The alternate approach, known as continuous projection, involves representing solutions using the same function space as FOMs and computing projection using continuous inner products. While continuous projection is more consistent with FOM, it is closely tied to the underlying discretization, making it challenging to enable nonintrusive model reduction \cite{Kalashnikova2014}.

\subsection{Proper orthogonal decomposition}
\label{sec:POD}

The first key ingredient of a successful reduced order modeling framework is selecting an adequate dimensionality reduction method. ROMs typically utilize high-fidelity simulation data to identify a suitable low-dimensional solution space that would minimize the degrees of freedom while having acceptable accuracy of solution representation. In this article, we consider POD, which is a linear or affine approach to reduce the dimensionality of the system. POD is the most popular approach and is widely applied to different applications. While POD may not be the most efficient dimensionality reduction approach for advection-dominated transport phenomenon due to a slow decay of Kolmogorov \textit{n}-width, the linear nature of POD allows lower offline stage cost and the possibility of mathematical analysis for error estimates and stability. Even though we will restrict the discussion of the nonintrusive model reduction approach proposed in this article to linear dimensionality reduction methods such as POD, the approach can be directly used in conjunction with nonlinear dimensionality reduction techniques such as autoencoders \cite{Lee2020, Kim2022}, neural fields \cite{Chen2023, Puri2024} and polynomial manifolds \cite{Barnett2022, Geelen2023}. 

POD is a linear dimensionality reduction method that approximates the solution as 
\begin{equation}
    \bm{u} \approx \bm{u}_0 + \sum_{i=1}^l a_i (t) \bm{\phi}_i ,
    \label{eq:PODvel}
\end{equation}
where $l$ is the total number of modes/states, $a_i : [0, T] \to \mathbb{R}$ is the $i^{th}$ temporally evolving reduced states, $\bm{\phi}_i \in \mathbb{R}^n$ is the $i^{th}$ spatial mode and $\bm{u}_0$ is the reference velocity, which is taken to be the zero vector in this article.  The total number of POD modes $l$ ultimately depends on the number of available timesteps of data. The set of spatial modes $\bm{\Phi}$ is the solution of the problem
\begin{equation}
    \bm{\Phi} = \underset{\hat{\bm{\Phi}} = \{ \bm{\phi}_1,\bm{\phi}_2,...,\bm{\phi}_{l} \}}{\text{arg min}} \sum_{j=1}^{l} \Big\vert \Big\vert \bm{u} - \sum_{i=1}^l a_i (t_j) \bm{\phi}_i \Big\vert \Big\vert^2_{L^2}
\end{equation}
such that $\bm{\phi}_i$ for $i = 1, 2, \cdot \cdot \cdot, \;l$ are orthonormal. These POD modes are the solution to the problem 
\begin{equation}
    \bm{\Phi} = \bm{X} \bm{\Psi} \sqrt{\bm{\Lambda}^{-1}},
\end{equation}
where $\bm{X} = [ \bm{u}(t_1) \; \bm{u}(t_2) \; \cdot \cdot \cdot \; \bm{u}(t_l)] \in \mathbb{R}^{n \times l}$ is the data matrix. The components of $\bm{\Psi} = [\bm{\psi}_1, \bm{\psi}_2, \cdot \cdot \cdot, \bm{\psi}_l]$ and $\bm{\Lambda} = \text{diag} (\lambda_1, \lambda_2, \cdot \cdot \cdot, \lambda_l)$ are obtained by solving the eigenproblem
\begin{equation}
    \bm{R} \bm{\psi}_i = \lambda_i \bm{\psi}_i,
\end{equation}
where $\bm{R} = \bm{X}^T \bm{X}$. For most scientific simulations, $l << n$, and therefore, a significant reduction of dimensionality is achieved. For common physical problems with a fast decay of Kolmogorov $n$-width, only the first $r << l$ energetic spatial modes need to be selected, thereby allowing the ROM degrees of freedom $r$ to be much smaller than FOM degrees of freedom $n$. 

\subsection{Galerkin and Petrov-Galerkin projections}
\label{sec:DiscreteProj}

The substitution of POD representation, shown in \eref{PODvel} with $\bm{u}_0$ as a zero vector, in discrete differential equations \eref{PDE_def} produces equations for the evolution of reduced states
\begin{equation}
    \sum_{i=1}^l \frac{d a_i}{d t} \bm{\phi_i} + \sum_{i=1}^l a_i \bm{L} \bm{\phi}_i + \bm{N} \Big( \sum_{i=1}^l a_i \bm{\phi}_i \Big) = 0,
    \label{eq:POD_PDE}
\end{equation}
which are $n$ equations for $l$ reduced states. The solution of this overdetermined system commonly involves the discrete projection of these equations to a low-dimensional solution subspace. This projection step follows ideas often borrowed from Galerkin and Petrov-Galerkin projections that are heavily used in finite element method literature \cite{Hughes2000}. Galerkin \cite{Aubry1988} and LSPG projections \cite{Carlberg2011, Carlberg2017} are the two most commonly used projection methods. A short description of other Petrov-Galerkin projections can be found in \cite{Parish2023}. 

Galerkin projection involves taking the inner product of \eref{POD_PDE} with $r$-dimensional linear subspace formed by $\bm{\phi}_k$ for $k \in \{1, 2, \cdot \cdot \cdot, r\}$ resulting in equations
\begin{equation}
    \frac{d a_k}{d t} + \sum_{i=1}^r a_i \bm{\phi}_k^T \bm{L} \bm{\phi}_i + \bm{\phi}_k^T \bm{N} \Big( \sum_{i=1}^r a_i \bm{\phi}_i \Big) = 0 \quad \forall \quad k \in \{1, 2, \cdot \cdot \cdot, r \}.
    \label{eq:ROM_Galerkin}
\end{equation}
This projection yields optimal ROM representation in continuous time, as shown in \cite{Carlberg2017}. With Galerkin projection on the subspace formed by $r$ most energy dominant states, the effect of unresolved $l - r$ least energy containing states is not accounted for. Therefore, Galerkin ROMs often have to be supplemented with closure terms \cite{Iliescu2014, Ahmed2021, Prakash2024a} to account for the effect of unresolved states on resolved states. We will not account for this effect in this article using the closure model. However, the current work also directly applies to these closure modeling approaches. In a common ROM scenario, $\bm{\phi}_i$ is obtained from high-fidelity simulation data, while $\bm{L}$ and $\bm{N}$ depend on underlying FOM discretization schemes. Therefore, by suitable selection of initial condition for $\bm{a}$, \eref{ROM_Galerkin} can be marched in time using suitable time integration schemes. Note that while the cost of evaluating $\bm{\phi}_k^T \bm{L} \bm{\phi}_i$ and $\bm{\phi}_k^T \bm{N} \Big( \sum_{i=1}^l a_i \bm{\phi}_i \Big)$ is $O(n)$, these terms can be precomputed in the offline stage of ROMs for most common PDEs. Therefore, the cost of the online stage, which involves integrating \eref{ROM_Galerkin} in time, is $O(r + p(r))$, where $p(r)$ is determined based on the nonlinearity of $\bm{N}$.

Another common projection method, referred to as LSPG, involves an oblique projection of \eref{POD_PDE} to a $r$-dimensional subspace formed by $\bm{\phi}_k$ for $k \in \{1, 2, \cdot \cdot \cdot r \}$. The projected equations are obtained by taking the inner product of \eref{POD_PDE} with the oblique projection vector $\bm{\psi}_k$ for $k \in \{1, 2, \cdot \cdot \cdot, r \}$ resulting in equations
\begin{equation}
    \sum_{i=1}^l \frac{d a_k}{d t} \bm{\psi}^T_k \bm{\phi}_i + \sum_{i=1}^l a_i \bm{\psi}_k^T \bm{L} \bm{\phi}_i + \bm{\psi}_k^T \bm{N} \Big( \sum_{i=1}^l a_i \bm{\phi}_i \Big) = 0 \quad \forall \quad k \in \{1, 2, \cdot \cdot \cdot, r \}.
    \label{eq:ROM_LSPG}
\end{equation}
\noindent The $r$-dimensional projection subspace is formed by vectors $\bm{\Psi} = [\bm{\psi}_1 \bm{\psi}_2 \cdot \cdot \cdot \bm{\psi}_r]$ that are obtained by solving the optimization problem
\begin{equation}
    \bm{\Psi} = \underset{\tilde{\bm{\Psi}}}{\text{arg min}} \Big\vert \Big\vert \tilde{\bm{\Psi}}^T \Big( \sum_{i=1}^l \frac{d a_i}{d t} \bm{\phi_i} + \sum_{i=1}^l a_i \bm{L} \bm{\phi}_i + \bm{N} \Big( \sum_{i=1}^l a_i \bm{\phi}_i \Big) \Big) \Big\vert \Big\vert_2^2.
    \label{eq:ROM_proj}
\end{equation}
The resulting equations from LSPG projection-based ROMs depend on the time integration scheme used for solving \eref{ROM_LSPG} and \eref{ROM_proj}. The selection of an explicit time-integration scheme implies that $\bm{\Psi} = \bm{\Phi}$. Therefore, these equations reduce to those for Galerkin projection-based ROMs, such as \eref{ROM_Galerkin}, as shown in \cite{Carlberg2017}. On the other hand, implicit time integration schemes are shown to yield optimal ROMs in discrete time \cite{Carlberg2017}. In this article, we use the backward Euler method for time integration for the equation of resolved states shown in \eref{POD_PDE}. In conjunction with this time integration scheme, the solution of \eref{POD_PDE} following LSPG projection reduces to the solution of the problem
\begin{equation}
    \bm{a}^{o+1} = \underset{\bm{z} = [z_1 z_2 \cdot \cdot \cdot z_r]^T}{\text{arg min}} \Big\vert \Big\vert \sum_{i=1}^r (z_i -a^{o}_i)\bm{\phi_i} + \Delta t \Big( \sum_{i=1}^r z_i \bm{L} \bm{\phi}_i + \bm{N} \Big( \sum_{i=1}^r z_i \bm{\phi}_i \Big)\Big) \Big\vert \Big\vert^2_2 
    \label{eq:LSPG_sol}
\end{equation}
where $\bm{a}^{o} \in \mathbb{R}^r$ is reduced state vector at the $o^{th}$ time step. This least squares problem is commonly solved using the Gauss-Newton method with hyper-reduction to reduce the online stage cost as in \cite{Carlberg2011}.

\subsection{Nonintrusive model reduction using S-LDOs}
\label{sec:SLDO-ROM}

In the previous section, we explored the development of Galerkin and LSPG projection-based ROMs. From \eref{ROM_Galerkin} and \eref{LSPG_sol}, we observe that both Galerkin and LSPG projection-based ROMs require the knowledge of differential operators $\bm{L}$ and $\bm{N}$. These differential operators are readily available in an intrusive ROM setting due to a suitable selection of discretization schemes and an appropriate supporting numerical framework of FOM to facilitate implementation. However, these differential operators are unknown in scenarios involving nonintrusive ROMs, as access to source code and discretization of FOMs may not be readily available. In such scenarios, these differential operators must be suitably selected to yield ROMs that respect physical properties, such as stability, which is typically carefully chosen for common discretization schemes yielding stable differential operators.

In the absence of high-fidelity differential operators $\bm{L}$ and $\bm{N}$, we can use high-fidelity simulation data to develop ROMs and determine modeled differential operators $\bm{L}^m$ and $\bm{N}^m$. In particular, we use S-LDOs as the modeled differential operators in this article. S-LDOs adequately represent high-fidelity differential operators by providing accurate and stable results for several linear and nonlinear PDE problems, as demonstrated through elaborate numerical experiments in \cite{Prakash2024b}. Therefore, S-LDOs are ideally suited to serve as a model for high-fidelity differential operators when these are unavailable, for example, in a nonintrusive setting. In such a scenario, the equation for reduced states of Galerkin projection-based ROMs is written as
\begin{equation}
    \frac{d a_k}{d t} + \sum_{i=1}^r a_i \bm{\phi}_k^T \bm{L}^m \bm{\phi}_i + \bm{\phi}_k^T \bm{N}^m \Big( \sum_{i=1}^r a_i \bm{\phi}_i \Big) = 0 \quad \forall \quad k \in \{1, 2, \cdot \cdot \cdot, r \},
    \label{eq:ROM_Galerkin_SLDO}
\end{equation}
where $\bm{L}^m$ and $\bm{N}^m$ are S-LDOs obtained by solving constrained regression problem in \eref{RegProbConstrainedFinalNonLin} and \eref{ConstrainedFinalNonLin}. Similarly, the evolution of reduced states for LSPG projection-based ROMs can be expressed as the solution to the problem
\begin{equation}
    \bm{a}^{o+1} = \underset{\bm{z} = [z_1 z_2 \cdot \cdot \cdot z_r]^T}{\text{arg min}} \Big\vert \Big\vert \sum_{i=1}^r (z_i -a^{o}_i)\bm{\phi_i} + \Delta t \Big( \sum_{i=1}^r z_i \bm{L}^m \bm{\phi}_i + \bm{N}^m \Big( \sum_{i=1}^r z_i \bm{\phi}_i \Big)\Big) \Big\vert \Big\vert^2_2 
    \label{eq:LSPG_sol_SLDO}
\end{equation}
where $\bm{a}^o \in \mathbb{R}^r$ are reduced states at the $o^{th}$ timestep. Note that the sparsity of S-LDOs ensures that the offline stage cost of determining ROMs is of the same order as that of intrusive ROMs. Furthermore, as the resulting equation form is the same as intrusive ROMs, the online stage cost is the same. Therefore, with just an additional step of obtaining S-LDOs from data in the offline stage provides us nonintrusive ROMs with similar computational costs as intrusive ROMs. 

Similar to the importance of stability while designing discretizations for high-fidelity simulations, stability for ROMs is also important. Stability is particularly important for dynamics forecasting problems where the ROMs are challenged to extrapolate in time. Several strategies have been proposed over the years to improve the stability characteristics of ROMs, with \cite{Rowley2004, Barone2009, Kalashnikova2010, Kalashnikova2014b, Amsallem2012, Carlberg2017, Iliescu2014} amongst just a few proposed methods. While most methods improve stability, there is limited work on mathematical proofs of guaranteed stability \cite{Rowley2004, Barone2009, Kalashnikova2010}. A comprehensive literature survey on ROM stabilization is given in \cite{Parish2023}. Using Theorem 1, we prove that S-LDOs can guarantee the stability of Galerkin projection-based ROMs.

\bigskip
\noindent \textbf{Theorem 1:} \textit{Consider a stable differential equation \eref{PDE_def_taylor} defined as per Lyapunov's indirect method, where the differential operator $\bm{N}_L$ is positive definite. The resulting ROM developed using POD and Galerkin projection is also stable following Lyapunov's indirect method.
}

\noindent \textbf{Proof:} Applying POD representation of solution shown in \eref{PODvel} to \eref{PDE_def_taylor}, we get
\begin{equation}
    \sum_{i=1}^l \frac{d a_i}{d t} \bm{\phi}_i + \sum_{i=1}^l a_i \bm{N}^L \bm{\phi}_i = 0.
\end{equation}
Following the Galerkin projection method of taking the inner product of this equation with $r$ energy dominant POD modes $\bm{\Phi} = [\bm{\phi}_1 \bm{\phi}_2 \cdot \cdot \cdot \bm{\phi}_r]$, we get
\begin{equation}
    \frac{d \bm{a}}{d t} + \bm{\bar{L}} \bm{a} = 0,
    \label{eq:Lyapu_deriv}
\end{equation}
where $\bm{\bar{L}} = \bm{\Phi}^T \bm{N}^L \bm{\Phi}$ and $\bm{a} = [a_1 a_2 \cdot \cdot \cdot a_r]^T$ is the resolved state vector and unresolved states $a_i$ for $i\in\{r+1, r+2, \cdot \cdot \cdot, l\}$ are neglected. For an arbitrary vector $\bm{x}$, 
\begin{equation}
    \bm{x}^T \bm{\bar{L}} \bm{x} = (\bm{\Phi} \bm{x})^T \bm{N}^L \bm{\Phi} \bm{x} \; > 0
\end{equation}
as $\bm{N}^L$ is positive definite, following stability based on Lyapunov's indirect method. As $\bm{\bar{L}}$ is positive definite, it has positive eigenvalues. Therefore, the resulting ROM obtained using POD and Galerkin projections is also stable. \qed

\bigskip
While we prove this theorem for the notion of stability following Lyapunov's indirect method, a similar result as holds for other notions of stability, such as energy stability, time stability, and Lyapunov's direct method, as shown in \cite{Rowley2004, Kalashnikova2014}. The solution to the regression problem in \eref{RegProbConstrainedFinalNonLin} and \eref{ConstrainedFinalNonLin} ensures $\bm{N}^L \succ 0$ for S-LDOs. Therefore, using Theorem 1, we conclude the resulting nonintrusive ROMs based on POD and Galerkin projection while using S-LDOs is also stable. 

\subsection{Computational Implementation}
\label{sec:Algo}

Computation of ROMs is typically divided into two stages: the expensive offline stage, which is typically performed only once, and the inexpensive online stage, which may be performed multiple times.
Using S-LDOs to enable nonintrusiveness in ROMs only involves an additional step in the offline stage of ROM development. The steps for the offline stage are given below:
\begin{enumerate}
    \item Obtain S-LDOs $\bm{L}^m$ and $\bm{N}^m$ from high-fidelity data of solution $\bm{u}$ and its time derivative following Algorithm 2 in \cite{Prakash2024b}.
    \item Determine POD modes $\bm{\Phi}$ using steps mentioned in Section \ref{sec:POD}.
    \item For Galerkin projection-based ROMs, ROM operators can be assembled offline for linear and nonlinear terms with polynomial nonlinearity.  
    \item After selecting a suitable time integration scheme, operators in LSPG projection-based ROMs may also be assembled in the offline stage. However, the process for determining these operators is typically more involved than that of Galerkin projection-based ROMs. 
\end{enumerate}

In the online stage of Galerkin projection-based ROMs, equations of reduced states in \eref{ROM_Galerkin_SLDO} are marched in time using a suitable time integrator while using hyperreduction if nonlinearity in \eref{ROM_Galerkin_SLDO} is not polynomial. For ROMs based on LSPG projection, \eref{LSPG_sol_SLDO} is directly solved using the Gauss-Newton method and a suitable hyperreduction scheme \cite{Carlberg2011} if the nonlinearity is not polynomial.

\subsection{Comparison to Operator Inference}
\label{sec:OpInfComp}

Another common strategy for nonintrusive model reduction is using operator inference \cite{Peherstorfer2016, Kramer2024} which follows \textit{project-then-learn} approach. To explain the difference between the proposed nonintrusive model reduction and operator inference approach, we consider \eref{ROM_Galerkin}
without the nonlinear term written in a vector form as
\begin{equation}
    \frac{d \bm{a}}{d t} + \bm{\Phi}^T \bm{L} \bm{\Phi} \bm{a} = 0.
    \label{eq:ROM_Galerkin_linear}
\end{equation}
As mentioned earlier, high-fidelity operators $\bm{L}$ are unknown in a nonintrusive setting. In operator inference, this equation is written as
\begin{equation}
    \frac{d \bm{a}}{d t} + \bm{L}^{\text{OpInf}} \bm{a}  = 0,    \label{eq:OpInf}
\end{equation}
where $\bm{L}^{\text{OpInf}}$ is the linear ROM operator learned from data by solving a regression problem \cite{Peherstorfer2016, Qian2022}. Therefore, operator inference directly determines the ROM operator $\bm{\Phi}^T \bm{L} \bm{\Phi}$ from data. Conversely, the approach proposed in our work determines $\bm{L}$ from data using S-LDOs and uses it to obtain ROM operator $\bm{\Phi}^T \bm{L} \bm{\Phi}$. While projection-based ROMs using both operator inference and S-LDOs yield nonintrusive ROMs, there are several differences in the capabilities of these two approaches as listed below.

\begin{itemize}
    \item \textbf{Computational cost:} The cost of the regression problem for operator inference is lower compared to obtaining S-LDOs. However, the cost of POD, which is required for both ROMs using operator inference and S-LDOs, is of similar order as the regression problem in S-LDOs. Therefore, the overall cost for the offline stage is not much different, although operator inference is less expensive.
    \item \textbf{Ease of integration to existing model reduction frameworks:} S-LDOs can be presented as a software package that could be integrated with existing intrusive ROM software frameworks. Furthermore, as demonstrated in Section \ref{sec:SLDO-ROM}, S-LDOs can be used with different stabilization methods. While this article demonstrates the applicability to Galerkin and LSPG projection-based ROMs, this method can be directly applied to other stabilization methods, such as Petrov-Galerkin projections \cite{Carlberg2017, Parish2023}, energetic inner product \cite{Rowley2004, Barone2009}, and eigenvalue reassignment \cite{Kalashnikova2014b, Rezaian2021}, and also to recently popular methods for nonlinear model-reduction methods \cite{Lee2020, Kim2022, Chen2023, Puri2024}. Conversely, the theory of operator inference directly involves Galerkin projection, and there are no direct pathways for application to other projections or stabilization methods. Therefore, operator inference acts as a standard-alone nonintrusive ROM approach.
    \item \textbf{Local spatial control for accuracy and stability:} As S-LDOs learn local differential operators, these operators can be tailored for a suitable tradeoff between accuracy and stability. For some scenarios, the stability constraint might be particularly strict and the accuracy of the regression problem is reduced. Having local spatial control on differential operator design allows us to force stability in certain regions of the problem domain while neglecting it in other regions. A suitable example of this requirement is the boundary layer in fluids, where strict stability requirements can significantly deplete the accuracy of predictions. Conversely, operator inference learns a global ROM operator without local spatial control over operator properties. Recent work on ensuring stability for operator inference \cite{Sawant2023, Tomoki2024} allows the applicability of operator inference for advection-dominated problems. However, these methods do not allow local spatial control over the stability properties.  
    
    \item \textbf{Complex multiphysics applications:} Operator inference is a fairly mature nonintrusive ROM method that has been demonstrated to be applicable for several multiphysics applications \cite{McQuarrie2021, Qian2022, Kramer2024}. As our article is the first to demonstrate S-LDOs for nonintrusive model reduction, additional work will be needed to extend its applicability to complex 3-D and multiphysics applications.
\end{itemize}

Using S-LDOs for nonintrusive ROMs offers an alternate strategy for nonintrusive model reduction as it exhibits several other positive attributes compared to operator inference. Therefore, the proposed approach can be taken instead of operator inference if the abovementioned scenarios are applicable. Operator inference is a very robust and mature approach for other common scenarios.

\section{Numerical results}
\label{sec:Results}

In this section, we demonstrate the applicability of nonintrusive Galerkin and LSPG projection-based ROMs using S-LDOs to determine ROM operators. We compare these results to intrusive Galerkin and LSPG projection-based ROMs, where underlying discretization is used as a differential operator to assemble equations of reduced states. This comparison is carried out for three PDE problems: 1) 1-D linear problem: Advection-diffusion equation, 2) 1-D nonlinear problem: Burgers equation, and 3) 2-D linear problem: Advection equation. Detailed numerical experiments validating S-LDOs for linear and nonlinear problems were conducted in \cite{Prakash2024b}. A stencil size of $11$, which was shown to yield optimal results for different cases, is selected. The constrained regression problem in S-LDOs is solved using sequential least squared programming optimizer available in \cite{2020SciPy} with the solver tolerance set to $10^{-10}$. The abbreviations for different ROM approaches are shown in \tabref{abbrv}.

\begin{table}
    \centering
    \caption{Abbreviatons for different ROM approaches.}
    \begin{tabular}{|c|c|}
        \hline
         \textbf{ROM approach} & \textbf{Abbreviation} \\
        \hline         
         Intrusive Galerkin & I-G \\
         Intrusive LSPG & I-LSPG \\
         Nonintrusive Galerkin using S-LDOs & NI-G-SLDO \\
         Nonintrusive LSPG using S-LDOs & NI-LSPG-SLDO \\
         \hline
    \end{tabular}
    \label{tab:abbrv}
\end{table}

\subsection{1-D scalar advection equation}

The first test case is the 1-D scalar advection equation
\begin{equation}
    \frac{\p u}{\p t} + c \frac{\p u}{\p x} = 0,
    \label{eq:1DadvecEq}
\end{equation}
where $c = 1.25 m/s$ is the advection velocity. 
To obtain FOM, the method of lines is used, which involves first spatially discretizing \eref{1DadvecEq} using first order-backward difference leading to a set of ODEs
\begin{equation}
    \frac{d \bm{u}}{d t} + \bm{L} \bm{u} = 0.
    \label{eq:1DadvecEq_ODE}
\end{equation}
The local differential equation for the $i^{th}$ degree of freedom is 
\begin{equation}
    \frac{d u_i}{d t} + \bm{L}^i \bm{u}_{\Omega^l_i} = 0,
    \label{eq:1DadvecEq_ODE_loc}
\end{equation}
where local linear operators $\bm{L}^i = \frac{c}{\Delta x}[-1 \; 1 \; 0]$, based on a local stencil of $\bm{u}_{\Omega^l_i} = [u_{i-1} \; u_i \; u_{i+1}]$, is assembled to form $\bm{L}$. The resulting equation in \eref{1DadvecEq_ODE} is integrated in time using the first-order forward Euler method to obtain FOM. The initial condition for this test case is chosen as $u(x,t = 0) = \exp{(-(x-3)^2)}$ for $x \in \Omega = [0,10]$. The equations of reduced states for this problem are obtained by using POD and Galerkin projection, resulting in the form
\begin{equation}
    \frac{d \bm{a}}{d t} + \bar{\bm{L}} \bm{a} = 0,
    \label{eq:ROM_1Dadvec}
\end{equation}
where $\bar{\bm{L}} = \bm{\Phi}^T \bm{L} \bm{\Phi}$ and $\bm{\Phi} = [\bm{\phi}_1 \bm{\phi}_2 \; \cdot \cdot \cdot \; \bm{\phi}_r]$. These equations are integrated in time using the same time integration scheme used for FOMs. For nonintrusive ROMs using S-LDOs, consistency with explicit time integration in FOMs is used to select appropriate time derivative and solution data in the respective regression problem. As explicit time integration is used for this case, LSPG and Galerkin projections are equivalent, and therefore, results for only Galerkin projection are presented. 

\begin{figure}[t]
    \centering
    \includegraphics[width=0.6\linewidth]{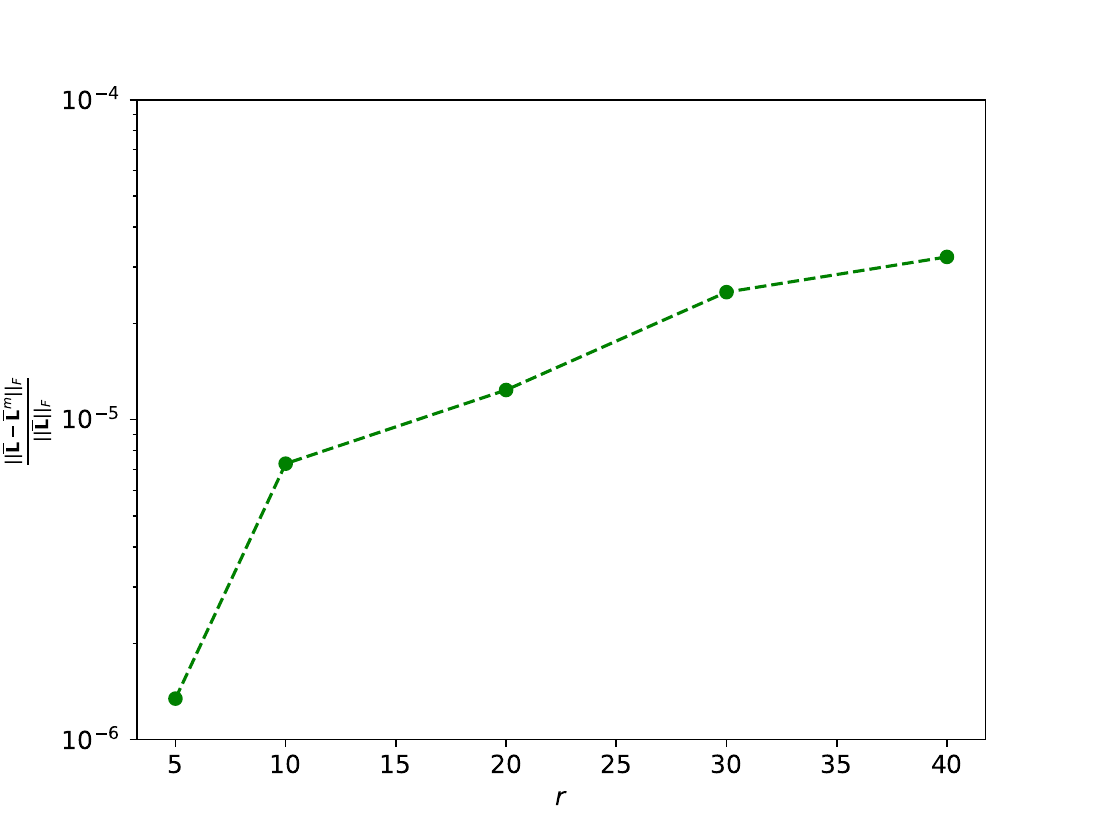}
    \caption{1-D scalar advection problem: Relative error in approximating the ROM operator using S-LDOs compared to intrusive Galerkin projection-based ROM.}
    \label{fig:OpError_advec}
\end{figure}

We first assess the closeness of ROM operators obtained for the nonintrusive Galerkin ROM using S-LDOs to the intrusive Galerkin ROM approaches. The proximity of these two operators is quantified using the error metric $\vert\vert \bar{\bm{L}} - \bar{\bm{L}}^m \vert \vert_F / \vert \vert \bar{\bm{L}} \vert \vert_F$, where $\bar{\bm{L}}$ is the intrusive ROM operator and $\bar{\bm{L}}^m$ is the nonintrusive ROM operator obtained using SLDOs. The comparison of error in approximating ROM operator for different numbers of modes is shown in \figref{OpError_advec}. The results indicate low errors in approximating the ROM operator for the proposed nonintrusive ROM approaches at fewer modes. This error gradually increases for a larger number of modes, but it remains low enough to yield accurate approximations of the ROM operator.

\begin{figure}
    \centering
    \subfigure[\label{fig:usol_t500_advec}]{\includegraphics[width=0.49\textwidth, trim={0.2cm 0cm 1.5cm 0.5cm},clip]{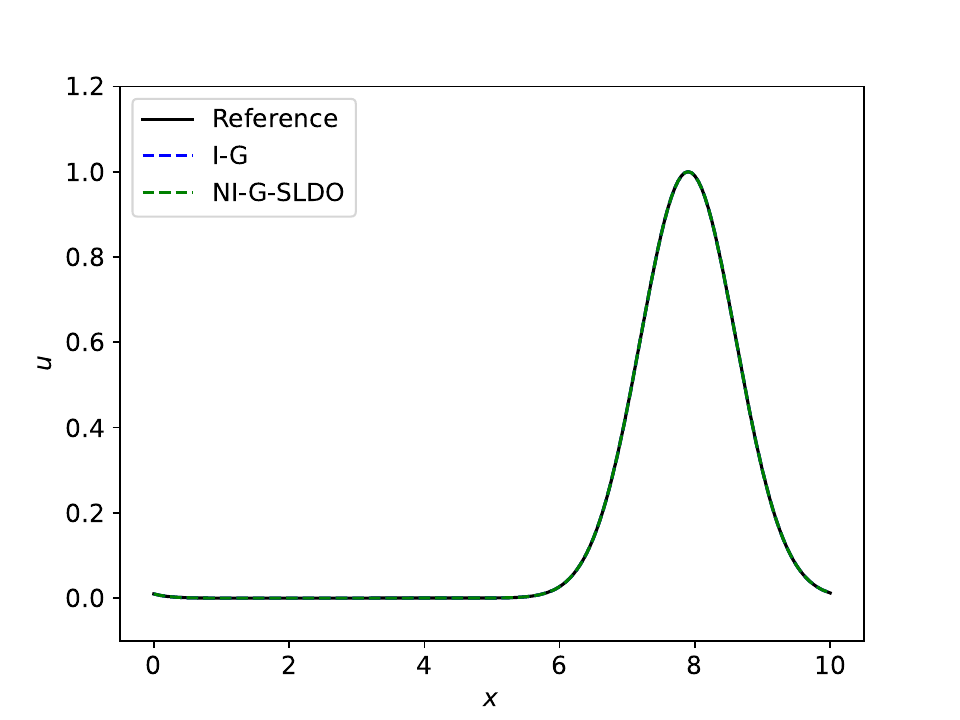}}
    \subfigure[\label{fig:usol_t5000_advec}]{\includegraphics[width=0.49\textwidth, trim={0.2cm 0cm 1.5cm 0.5cm},clip]{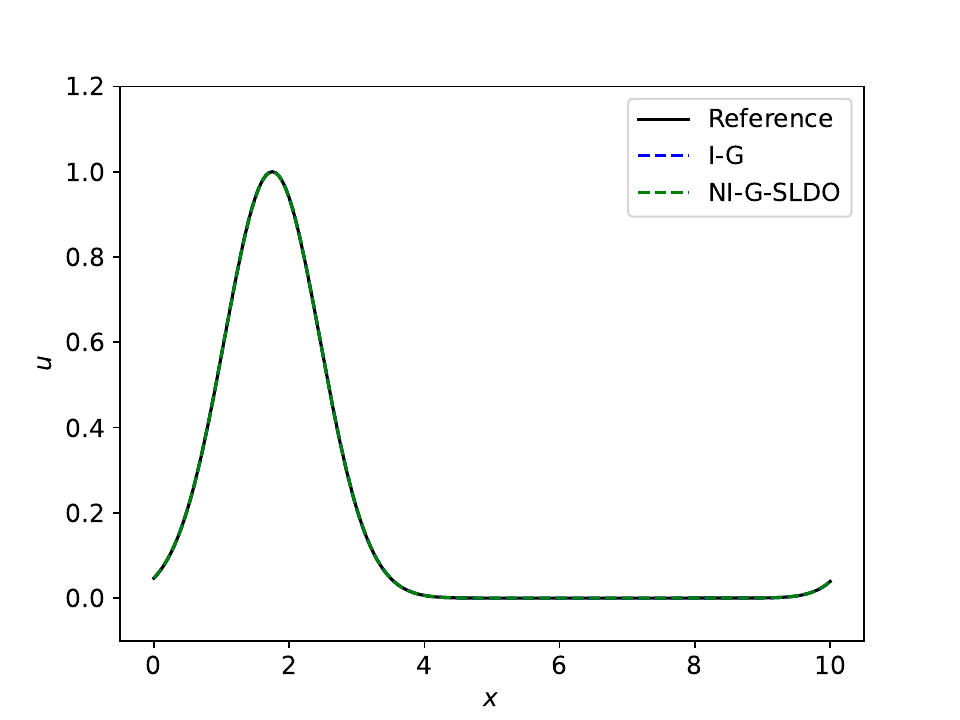}}
    \vspace{-3mm}
    \caption{1-D scalar advection problem: Solution prediction with different ROMs for $r = 20$ modes at (a) $t = 20s$ and (b) $t = 200s$. All the tested ROMs have overlapping solutions.}
    \label{fig:usol_1Dadvec}
\end{figure}

The predicted solution obtained using different ROM approaches with $r = 20$ modes is compared to FOM results at two different time instances in \figref{usol_1Dadvec}. Nonintrusive ROMs using S-LDOs produce visually indistinguishable solutions compared to intrusive ROM and FOM predictions. This behavior holds for both time instances, at $t = 20s$, which is included in the dataset used to obtain ROMs, and $t = 200s$, which is a future time not included in the training dataset.

In addition to comparing the solution to the reference FOM results, we assess the variation of relative error in time
\begin{equation}
    e_u (t) = \frac{\vert \vert \bm{u} (t) - \bm{u}^m (t) \vert \vert_2^2 }{\vert \vert \bm{u} (t) \vert \vert_2^2},
    \label{eq:error_u}
\end{equation}
where $\bm{u}$ is the FOM solution and $\bm{u}^m$ is the solution predicted using ROMs. We also define a second error metric called total relative error, which is obtained by taking the $L_2$ norm in space and time, which corresponds to the Frobenius norm $\vert\vert \cdot \vert \vert_F$ if the solution is stored as a 2-D matrix. This error metric is written as
\begin{equation}
    \varepsilon_{xt} = \frac{\vert \vert \bm{u} (\cdot) - \bm{u}^m (\cdot) \vert \vert_F }{\vert \vert \bm{u} (\cdot) \vert \vert_F}
    \label{eq:def_totalerror}
\end{equation}
and compared for different ROMs at several selections of reduced states.  

\begin{figure}
    \centering
    \subfigure[\label{fig:errorTime_r10}]{\includegraphics[width=0.32\textwidth, trim={0.2cm 0cm 1.5cm 0.5cm},clip]{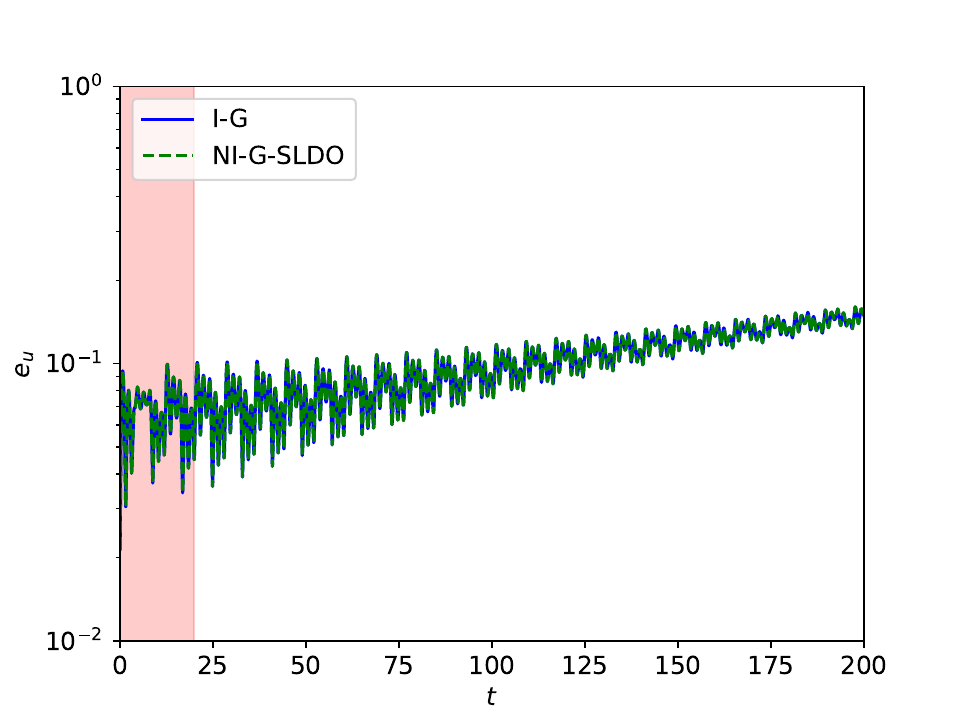}}
    \subfigure[\label{fig:errorTime_r40}]{\includegraphics[width=0.32\textwidth, trim={0.2cm 0cm 1.5cm 0.5cm},clip]{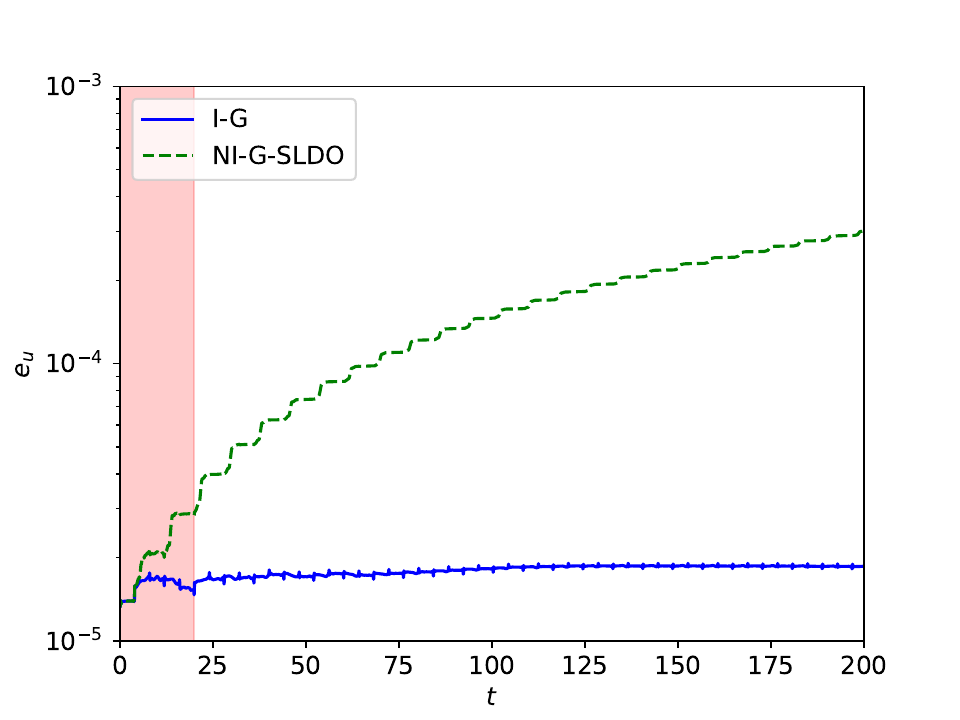}}
    \subfigure[\label{fig:Totalerror_advec}]{\includegraphics[width=0.32\textwidth, trim={0.2cm 0cm 1.5cm 0.5cm},clip]{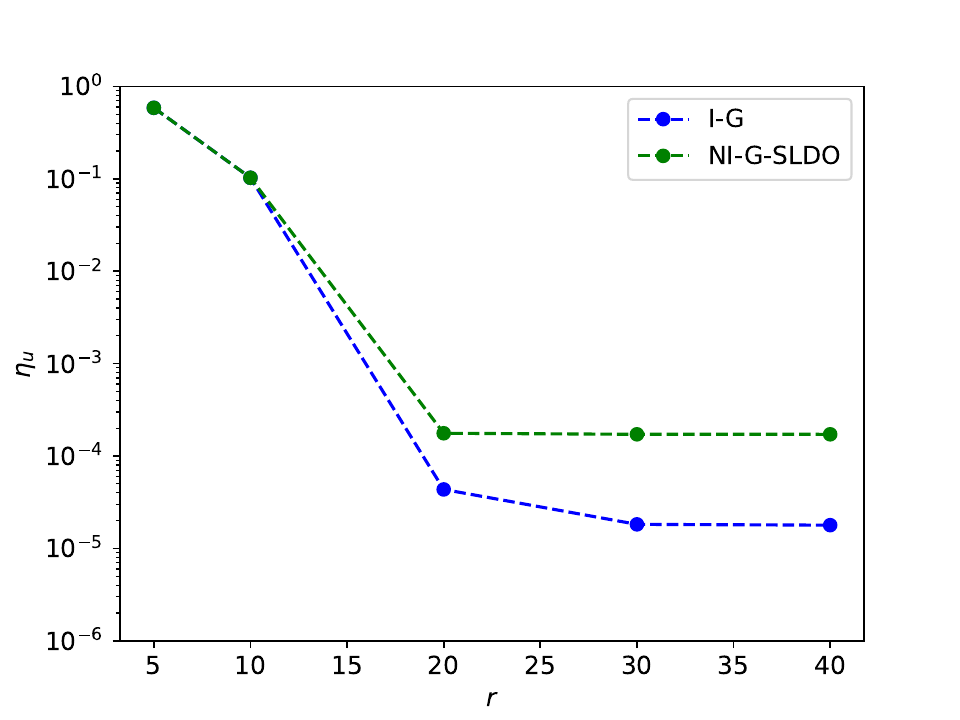}}    
    \vspace{-3mm}
    \caption{1-D scalar advection problem: Variation of relative error in predicted solution in time for different ROMs with number of modes (a) $r = 10$ and (b) $r = 40$. (c) Total relative error in space and time for different mode selections 
    The red-shaded region in (a) and (b) marks when the data is extracted to learn ROMs. The unshaded region is the time of extrapolation for ROMs.}
    \label{fig:errorTime_1Dadvec}
\end{figure}

The variation of relative error in time for different ROM approaches is shown in \figref{errorTime_r10} and \figref{errorTime_r40}. These results indicate that errors in predicted solutions are similar for nonintrusive and intrusive methods for $r = 10$ modes. This error remains low outside the initial time interval used to determine ROMs. The error for all ROM approaches decreases for a larger number of modes, and differences between intrusive and nonintrusive ROMs become more apparent. For $r = 40$ modes, intrusive ROMs exhibit very low errors that are consistently low even outside the initial time window used to determine ROMs. On the other hand, nonintrusive ROMs exhibit an increase in error by an order of magnitude outside the initial time window. Despite this increase in error, the overall errors still remain very low, demonstrating the feasibility of using nonintrusive ROMs. The variation of total error with different selections of resolved modes is shown in \figref{Totalerror_advec}. These results indicate that both nonintrusive and intrusive ROMs yield similar results for the lower number of modes. On the contrary, nonintrusive methods are less accurate at a higher number of modes as they exhibit higher total error. 

This test case indicated that different projection-based ROM approaches give accurate results for hyperbolic problems such as the 1-D advection equation not only for the initial time window used to obtain ROMs but also extrapolate well with low errors for dynamics forecasting. Nonintrusive ROMs using SLDOs provide comparably good results to intrusive ROMs, especially at fewer modes. This behavior is particularly encouraging as ROMs should ideally have a lower number of modes, as ROMs with a higher number of modes will have higher online stage costs, which would discourage using ROMs. Both intrusive and nonintrusive ROMs require a large number of modes to yield very low errors. This behavior is due to the slow decay of Kolmogorov $n$-width exhibited by this advection-dominated problem. Alternate approaches for nonlinear model reduction, such as those proposed in \cite{Lee2020, Kim2022, Barnett2022}, can be directly combined with S-LDOs to yield nonintrusive nonlinear ROMs, such as \cite{Geelen2023}, that are highly desirable for advection-dominated transport problems.

\subsection{1-D Burgers equation}

For the second numerical experiment, we assess the performance of different ROM approaches for 1-D Burgers equation, which is a nonlinear PDE of the form
\begin{equation}
    \frac{\p u}{\p t} + u \frac{\p u}{\p x} = \nu \frac{\partial^2 u}{\p x^2},
    \label{eq:1DBurgersEq}
\end{equation}
where $\nu$ is the viscosity parameter that governs the diffusion term. The FOM is obtained by following the method of lines where the spatial discretization is first order backward difference for the advection term and second order centered difference for the diffusion term. The resulting system of ODEs is
\begin{equation}
    \frac{d \bm{u}}{d t} + \bm{N} \bm{z}(\bm{u}) + \bm{L} \bm{u} = 0,
    \label{eq:1DBurgersEq_ODE}
\end{equation}
where $\bm{N}$ and $\bm{L}$ are nonlinear differential operators. The equations for the $i^{th}$ degree of freedom are
\begin{equation}
    \frac{d u_i}{d t} + \bm{N}^i \bm{z}(\bm{u}_{\Omega^n_i}) + \bm{L}^i \bm{u}_{\Omega^l_i} = 0,
    \label{eq:1DBurgersEq_ODE_loc}
\end{equation}
where local nonlinear operator $\bm{N}^i = \frac{1}{\Delta x} [-1 \; 1 \; 0]$ based on the nonlinear stencil $\bm{z} (\bm{u}_{\Omega^n_i}) = u_i [u_{i-1} \; u_i \; u_{i+1}]^T$ and local linear operators $\bm{L}^i = \frac{\nu}{(\Delta x)^2} [-1 \; 0 \; 1]$ based on the linear stencil $\bm{u}_{\Omega^l_i} = [u_{i-1} \; u_i \; u_{i+1}]^T$ are assembled to obtain  $\bm{N}$ and $\bm{L}$. The initial condition for this test case is chosen to be $u(x,t = 0) = \sin{(x)}$ for $x \in \Omega = [0,2\pi]$. For this validation case, we select $\nu = 10^{-3}$ with results in the formation of shock at $t = 3s$. In such scenarios, an implicit time integration is favorable as it leads to a substantial reduction in computational cost by enabling the selection of larger timesteps. Therefore, we use first order backward Euler method for integrating  \eref{1DBurgersEq_ODE} in time. For Galerkin ROMs, equations of reduced states are obtained by using POD and Galerkin projection as shown for the previous numerical experiment. We ensure consistency in time marching between FOMs and ROMs by using the Euler backward method for all ROM approaches. The selection of implicit time integration for ROMs also enables the comparison of ROMs using S-LDOs with both Galerkin and LSPG projections.  

\begin{figure}[t]
    \centering
    \includegraphics[width=0.6\linewidth]{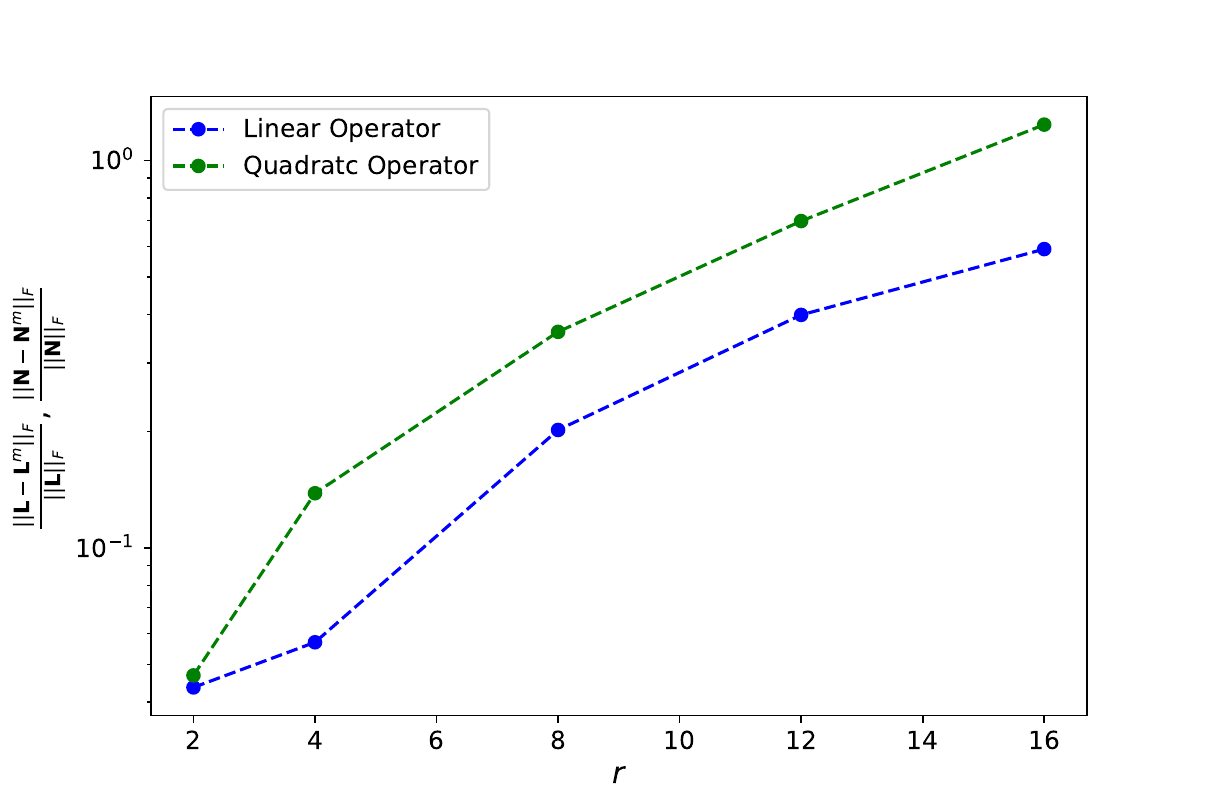}
    \caption{1-D Burgers problem: Relative error in approximating the ROM operator using S-LDOs compared to intrusive Galerkin projection-based ROM.}
    \label{fig:OpError_Burger}
\end{figure}

For Galerkin ROMs, we first quantify the relative error in approximating the linear operator, defined as $\vert \vert \bar{\bm{L}} - \bar{\bm{L}}^m \vert \vert_F / \vert \vert \bar{\bm{L}} \vert \vert_F$, and the nonlinear operator, defined as $\vert \vert \bar{\bm{N}} - \bar{\bm{N}}^m \vert \vert_F / \vert \vert \bar{\bm{N}} \vert \vert_F$, where the superscript $m$ indicates the operator obtained using S-LDOs. These errors are compared for different numbers of modes in \figref{OpError_Burger}. The magnitude of these errors is higher than that of the advection problem. However, these still remain low, especially at a lower number of modes. We observe a gradual increase in these errors with an increase in the number of modes indicating reduced accuracy for these higher modes. The errors for both linear and quadratic operators follow a similar trend, with quadratic operators having higher errors at a higher number of modes. This result indicates that reduced accuracy at higher mode selection has a more pronounced effect on the nonlinear terms.

\begin{figure}[t]
    \centering
    \subfigure[\label{fig:usol_t500_Burger_true}]{\includegraphics[width=0.28\textwidth, trim={0.5cm 0cm 1.5cm 0.5cm},clip]{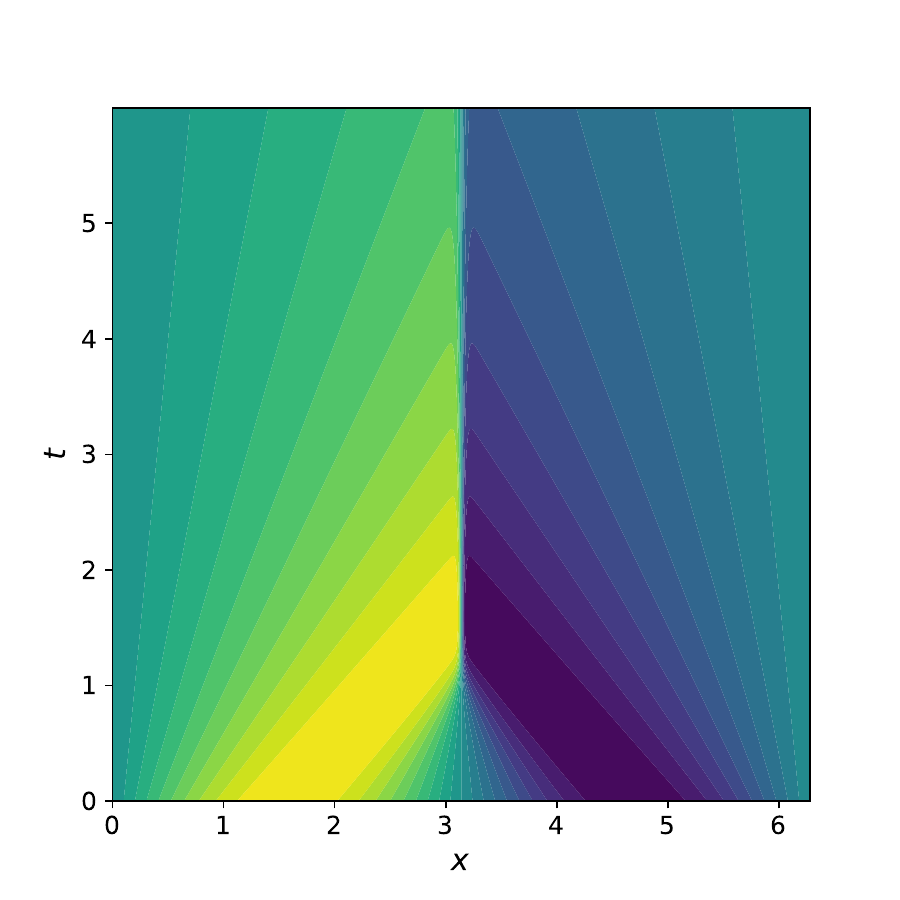}}
    \includegraphics[width=0.062\textwidth, trim={11.8cm 0cm 0.5cm 1cm},clip]{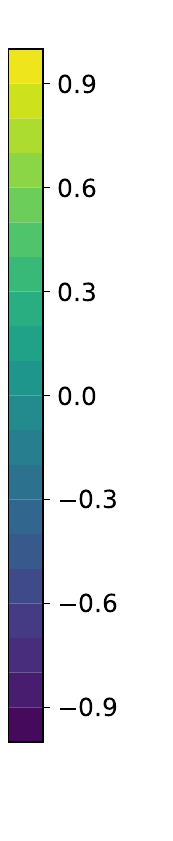}  
    \subfigure[\label{fig:usol_t500_Burger_IG}]{\includegraphics[width=0.28\textwidth, trim={0.5cm 0cm 1.5cm 0.5cm},clip]{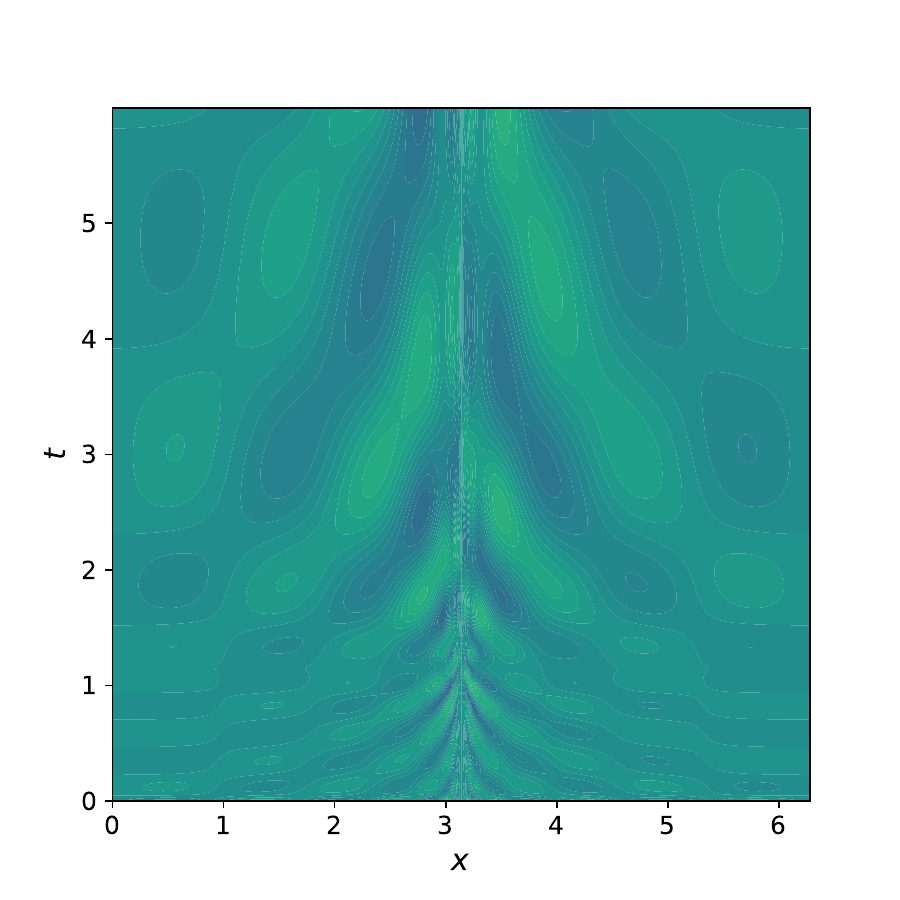}}
    \subfigure[\label{fig:usol_t500_Burger_NIG}]{\includegraphics[width=0.28\textwidth, trim={0.5cm 0cm 1.5cm 0.5cm},clip]{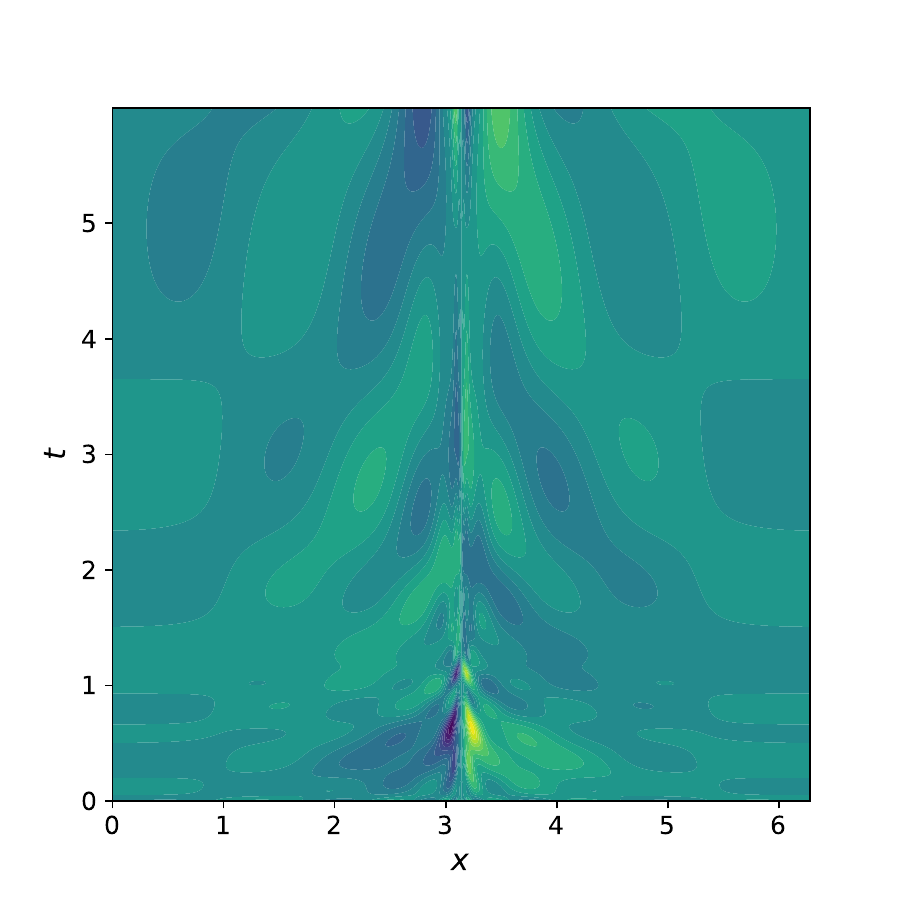}}
    \includegraphics[width=0.062\textwidth, trim={11.8cm 0cm 0.5cm 1cm},clip]{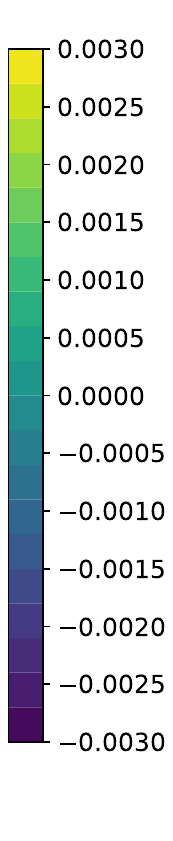}  
    \vspace{-3mm}
    \caption{1-D Burgers problem: (a) Spatio-temporal solution prediction for the reference FOM. Errors ($u^m (x,t) - u (x,t)$) in spatio-temporal solution prediction for (b) I-G ROM and (c) NI-G-SLDO ROM for $r = 8$ modes.}
    \label{fig:usol_1DBurgers}
\end{figure}

\begin{figure}
    \centering
    \subfigure[\label{fig:errorTime_r4_Burger}]{\includegraphics[width=0.32\textwidth, trim={0.2cm 0cm 1.5cm 0.5cm},clip]{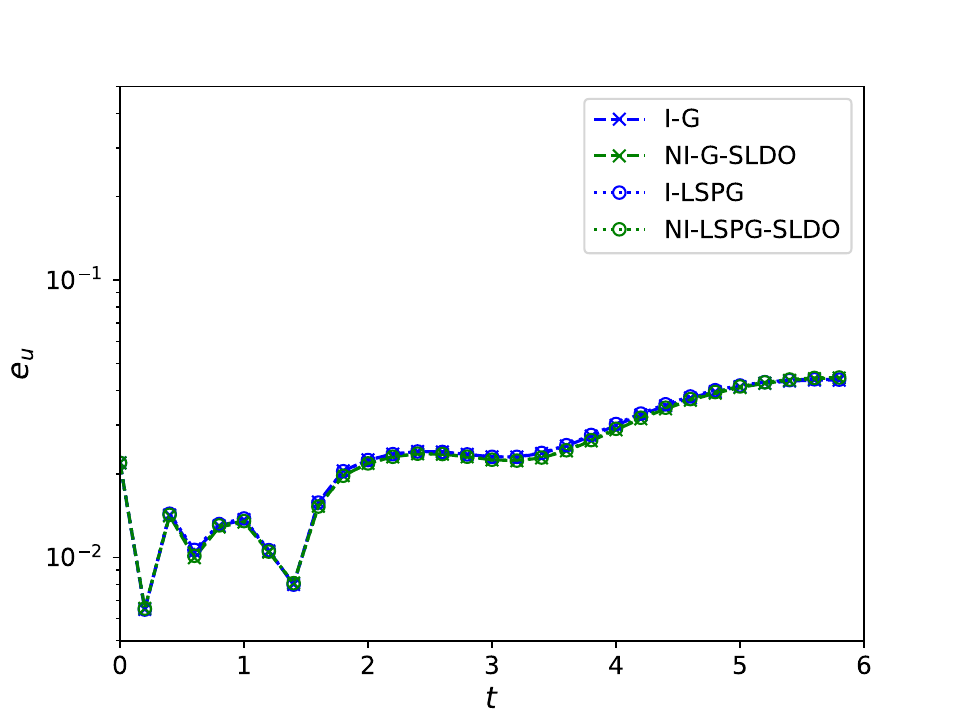}}
    \subfigure[\label{fig:errorTime_r12_Burger}]{\includegraphics[width=0.32\textwidth, trim={0.2cm 0cm 1.5cm 0.5cm},clip]{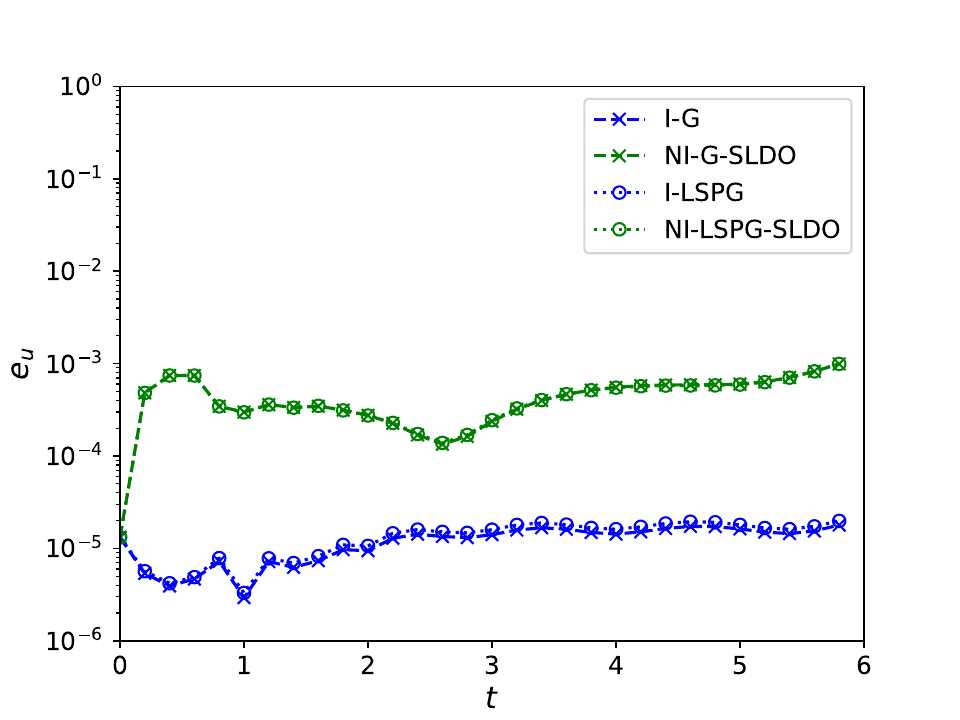}}
    \subfigure[\label{fig:Totalerror_advec_Burger}]{\includegraphics[width=0.32\textwidth, trim={0.2cm 0cm 1.5cm 0.5cm},clip]{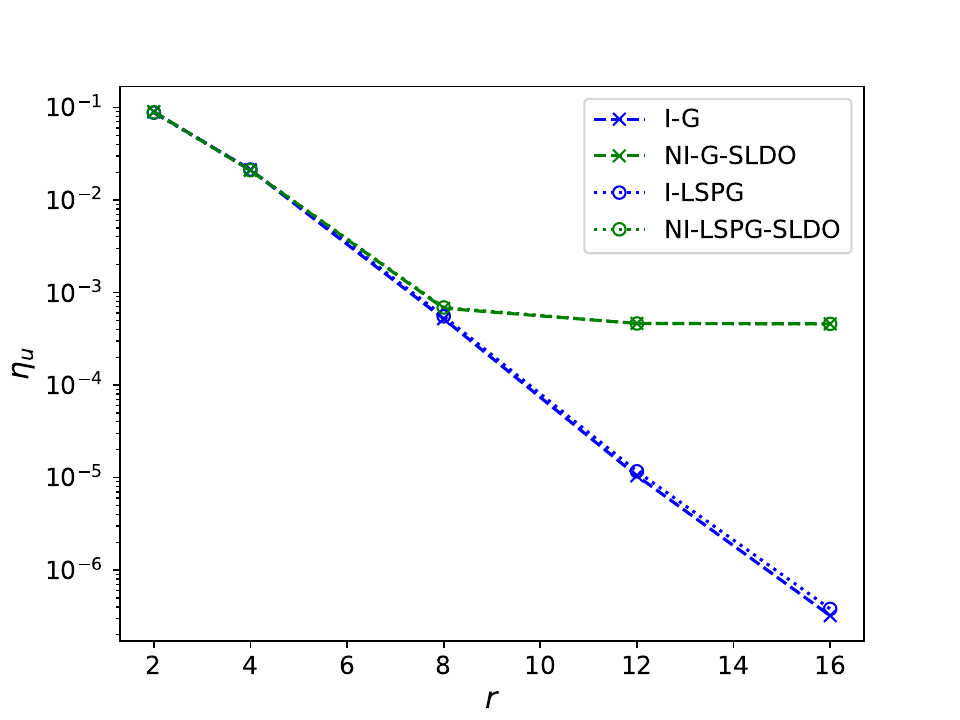}}    
    \vspace{-3mm}
    \caption{1-D Burgers problem: Variation of relative error in predicted solution in time for different ROMs with $r$ number of modes for (a) $r = 4$ and (b) $r = 12$. (c) Total relative error in space and time for different mode selections.}
    \label{fig:errorTime_1DBurgers}
\end{figure}

The evolution of the error in solution prediction in time for different ROMs is shown in \figref{usol_1DBurgers}. The predicted solution using intrusive and nonintrusive Galerkin projection-based ROMs exhibits a very low error with respect to the reference FOM solution. The accuracy of these predictions is further quantified by assessing error metrics defined in \eref{error_u} and \eref{def_totalerror}. The variation of relative error in time for different ROMs is shown in \figref{errorTime_r4_Burger} and \figref{errorTime_r12_Burger} for $r = 4$ and $12$ modes, respectively. The results indicate that errors for both intrusive and nonintrusive ROMs are similar for $r = 4$ modes. Conversely, the errors for nonintrusive ROMs are much larger (by an order of magnitude) than those for intrusive ROMs for $r = 12$ modes. The variation of total error for different number of modes is shown in \figref{Totalerror_advec_Burger}. These results indicate that both intrusive and nonintrusive ROMs exhibit very similar errors at a lower number of modes. However, at a higher number of modes, the errors for nonintrusive ROMs asymptote to a value, while the errors for intrusive ROMs continue decreasing with additional modes. The results also indicate that Galerkin and LSPG projections yield similar ROM performance for this problem, as also observed in prior studies \cite{Kim2022}. For more complex nonlinear PDEs, LSPG projection-based ROMs have been shown to yield much better results than its Galerkin counterpart \cite{Carlberg2017, Grimberg2020}. The main goal of this article is to demonstrate that S-LDOs can enable nonintrusive model reduction without a significant reduction in accuracy compared to the intrusive counterpart. The results validate this observation, which shows similar errors between intrusive and nonintrusive ROMs for both Galerkin and LSPG projection-based ROMs.

This problem demonstrates the success of S-LDOs in enabling nonintrusive model reduction for nonlinear PDEs. For a lower number of modes, which are desired in common model reduction scenarios, nonintrusive ROMs using S-LDOs exhibit very similar accuracy compared to intrusive ROMs. The asymptotic behavior of nonintrusive ROMs for a higher number of modes is due to a reduction in the consistency of approximating differential operators using S-LDOs. For an even more significant number of modes, which are not shown here, the errors for intrusive ROMs also asymptote to a value. These results indicate the intrusive ROMs are more consistent than the proposed nonintrusive ROMs. However, this improved consistency comes at the cost of intrusiveness which is undesirable for several applications. The good performance of nonintrusive LSPG projection-based ROMs highlights the success of S-LDOs in enabling nonintrusive LSPG projection-based ROMs. While we demonstrate this applicability using LSPG projections as an example, S-LDOs can directly enable nonintrusive model reduction for other commonly used Petrov-Galerkin projections \cite{Parish2023}. 

\subsection{2-D scalar advection equation}

For the third numerical experiment, we consider a 2-D scalar advection equation. This validation case demonstrates the applicability of the proposed nonintrusive ROM approach for 2-D problems with slow decay of Kolmogorov $n$-width. This equation is given as
\begin{equation}
    \frac{\p u}{\p t} + \bm{c} \cdot \nabla u = 0,
\end{equation}
where $\bm{c} = [c_x \; c_y]^T$ is advection velocity vector. FOMs are obtained by using first order backward difference for discretization in space, which leads to the equation
\begin{equation}
    \frac{d \bm{u}}{d t} + \bm{L}_x \bm{u} + \bm{L}_y \bm{u} = 0,
    \label{eq:2Dadvec_ODE}
\end{equation}
where $\bm{L}_x$ and $\bm{L}_y$ are linear differential operators that correspond to the two cartesian spatial gradients. The equation for the $i^{th}$ degree of freedom is
\begin{equation}
    \frac{d u_i}{d t} + \bm{L}^i_x \bm{u}_{\Omega^x_i} +  \bm{L}^i_y \bm{u}_{\Omega^y_i} = 0,
    \label{eq:2Dadvec_Eq}
\end{equation}
where $\bm{L}^i_x$ and $\bm{L}^i_y$ are local linear differential operators, while $\bm{u}_{\Omega^x_i}$ and $\bm{u}_{\Omega^y_i}$ are local solution stencils corresponding to these differential operators. These local differential operators $\bm{L}^i_x$ and $\bm{L}^i_y$ are assembled to form differential operators $\bm{L}_x$ and $\bm{L}_y$ in \eref{2Dadvec_ODE}. For FOMs, we use first order forward Euler explicit time integration method to march \eref{2Dadvec_ODE} in time. Equations of reduced states, similar to \eref{ROM_1Dadvec}, are obtained using POD representation of the solution in \eref{2Dadvec_Eq} and applying Galerkin projection. These equations are integrated in time using the same time integration schemes as the FOM. Note that the selected combination of spatial and temporal discretizations leads to an over-diffusive FOM due to artificial diffusion terms arising as truncation errors of the chosen discretization schemes. Even though such effects are not desirable in a FOM, the assessment of nonintrusive ROMs for such scenarios validates the ability of ROMs to give consistent solutions to the underlying FOM. 

\begin{figure}[t]
    \centering
    \subfigure[\label{fig:usol_t400_2DAdvec_true}]{\includegraphics[width=0.28\textwidth, trim={0.5cm 0cm 1.5cm 0.5cm},clip]{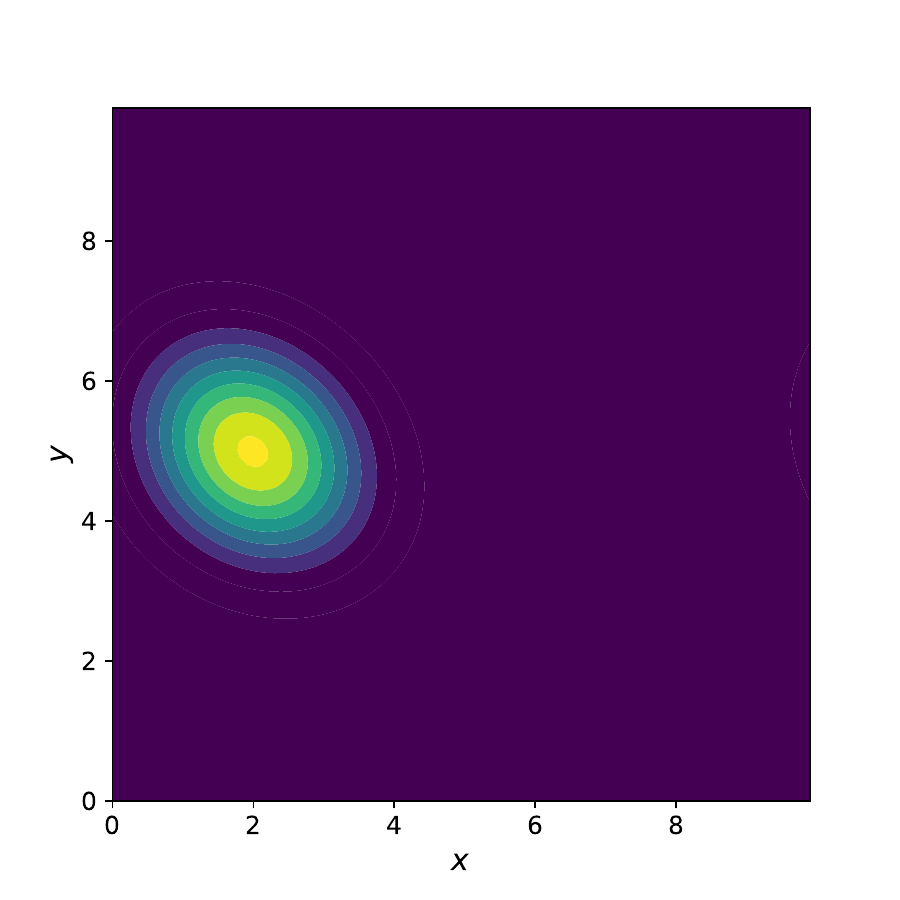}}
    \includegraphics[width=0.059\textwidth, trim={11.8cm 0cm 0.65cm 1cm},clip]{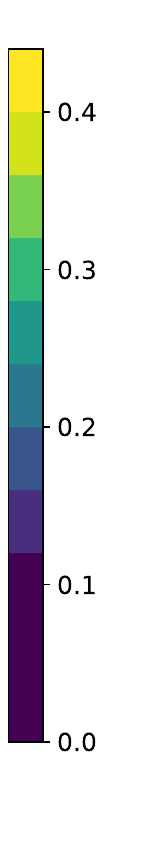}  
    \subfigure[\label{fig:usol_t400_2DAdvec_IG}]{\includegraphics[width=0.28\textwidth, trim={0.5cm 0cm 1.5cm 0.5cm},clip]{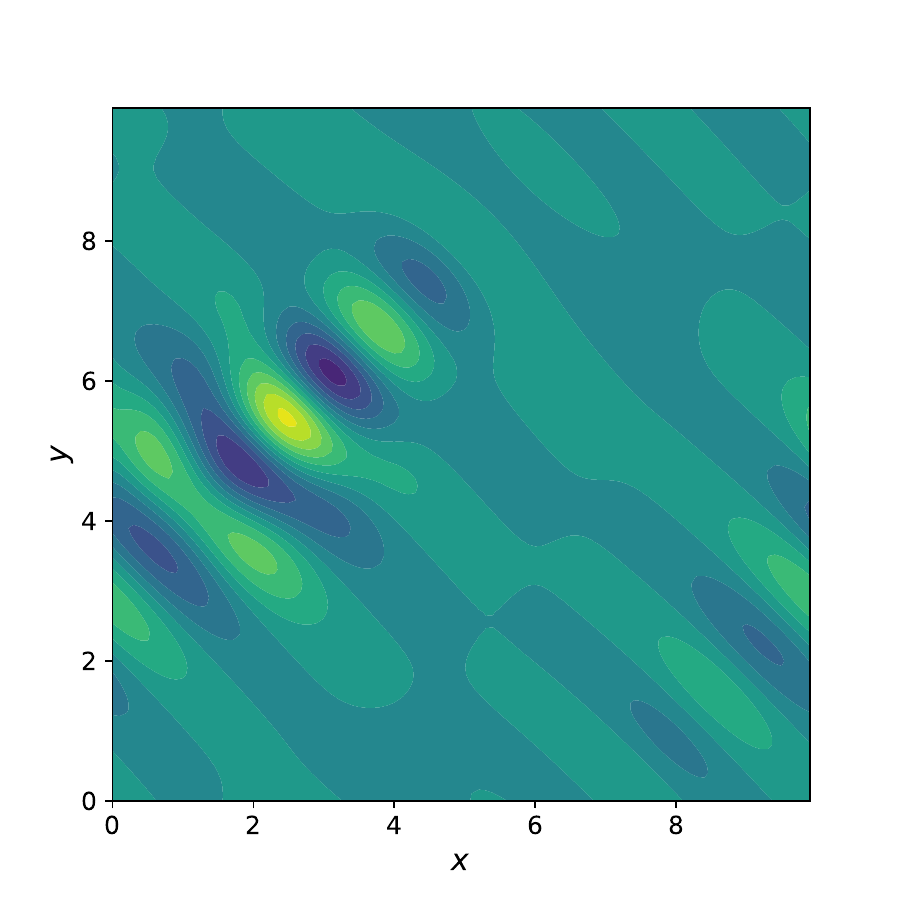}}
    \subfigure[\label{fig:usol_t400_2DAdvec_NIG}]{\includegraphics[width=0.28\textwidth, trim={0.5cm 0cm 1.5cm 0.5cm},clip]{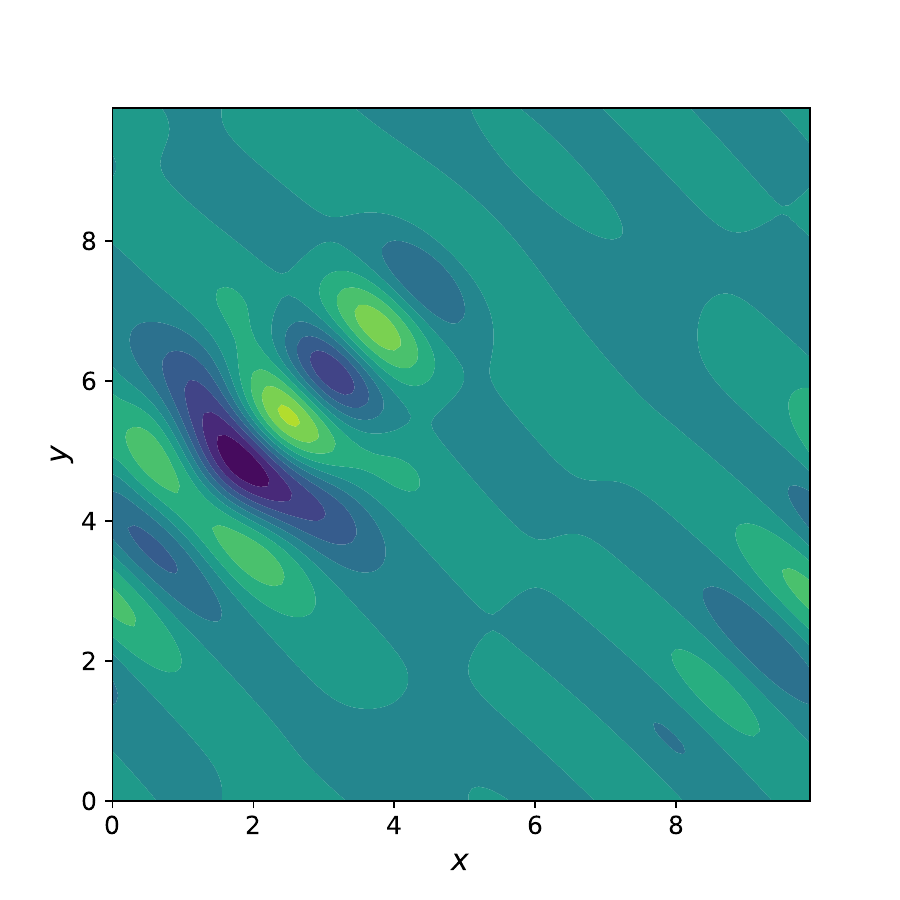}}
    \includegraphics[width=0.059\textwidth, trim={11.8cm 0cm 0.65cm 1cm},clip]{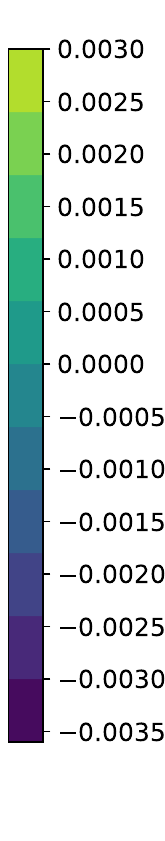}  
    \vspace{-3mm}
    \caption{2-D advection problem: (a) Reference FOM solution prediction at $t = 20s$. Error in solution ($u^m (x,y) - u (x,y)$) at $t = 20s$ predicted using (b) I-G ROM and (c) NI-G-SLDO ROM for $r = 20$ modes.}
    \label{fig:usol_2DAdvec}
\end{figure}

\begin{figure}
    \centering
    \subfigure[\label{fig:errorTime_r20_2Dadvec}]{\includegraphics[width=0.32\textwidth, trim={0.2cm 0cm 1.5cm 0.5cm},clip]{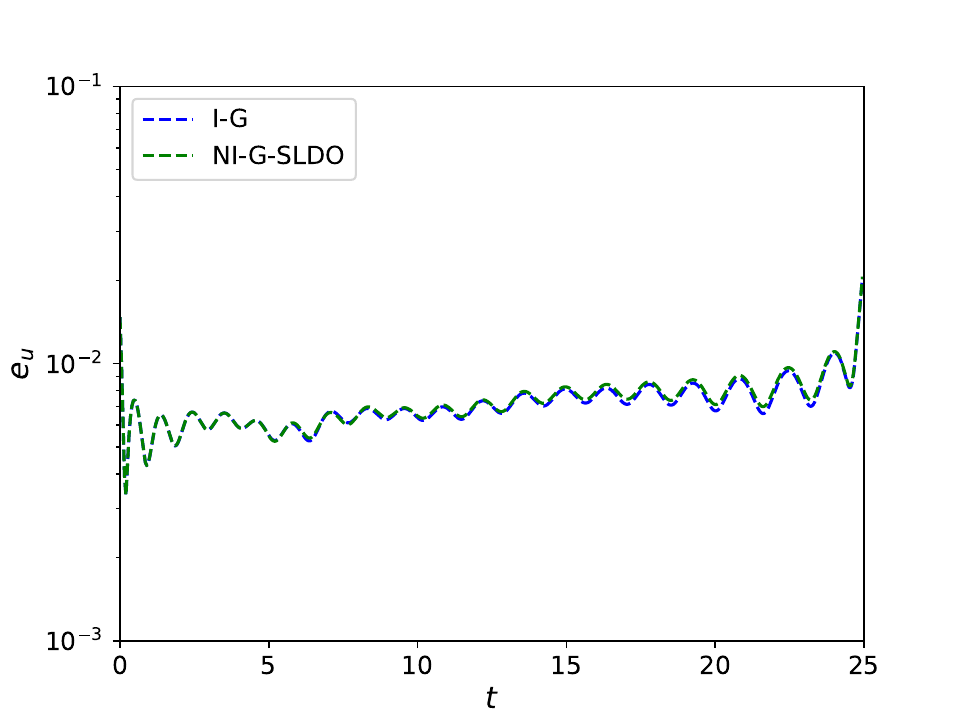}}
    \subfigure[\label{fig:errorTime_r30_2Dadvec}]{\includegraphics[width=0.32\textwidth, trim={0.2cm 0cm 1.5cm 0.5cm},clip]{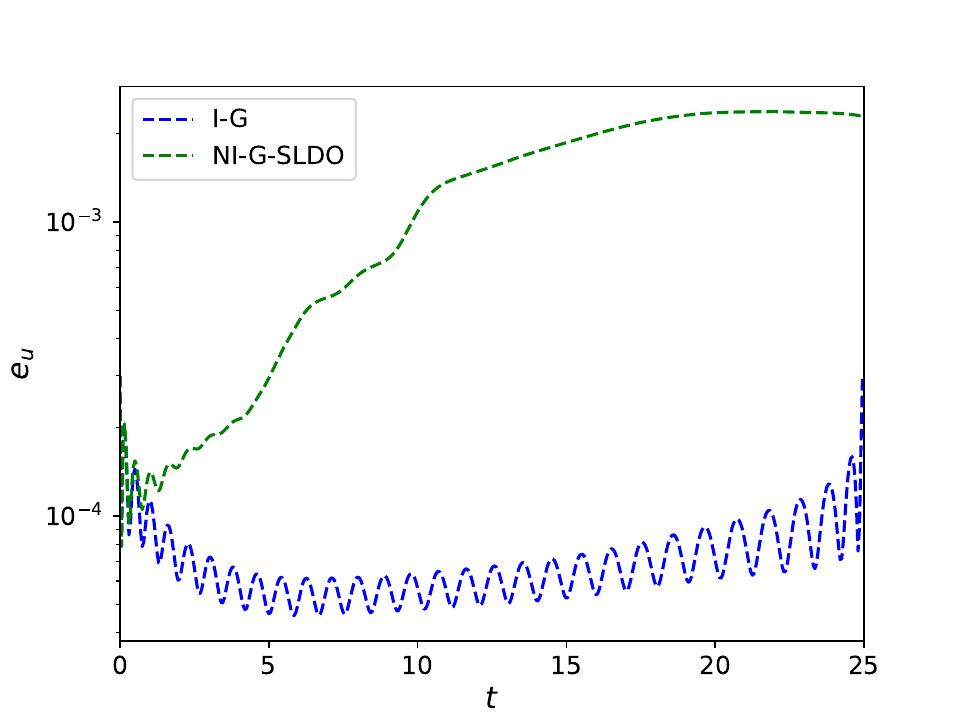}}
    \subfigure[\label{fig:Totalerror_2Dadvec}]{\includegraphics[width=0.32\textwidth, trim={0.2cm 0cm 1.5cm 0.5cm},clip]{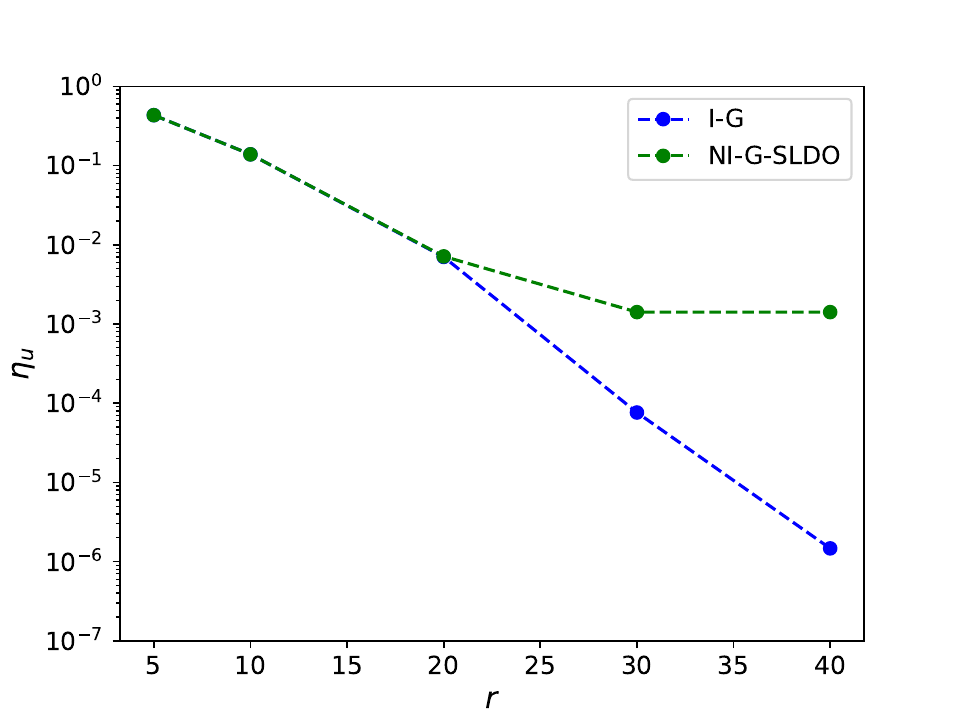}}    
    \vspace{-3mm}
    \caption{2-D advection problem: Variation of relative error in predicted solution in time for different ROMs with $r$ number of modes for (a) $r = 20$ and (b) $r = 30$. (c) Total relative error in space and time for different mode selections.}
    \label{fig:errorTime_2DAdvec}
\end{figure}

The errors between S-LDO enabled nonintrusive ROM operators, and intrusive ROM operators were observed to behave similarly to the previous two numerical experiments. Therefore, we have excluded this discussion for brevity. Instead, we directly assess the prediction of the solution obtained by integrating equations of reduced states in time. The errors in the predicted solution obtained using different ROMs are compared at $t = 20s$ in \figref{usol_2DAdvec}. The predicted solutions using intrusive and nonintrusive ROMs agree well with FOM predictions and exhibit low errors. We observe a similar error distribution for intrusive and proposed non-intrusive ROMs. A more comprehensive comparison is conducted by plotting relative errors for intrusive and nonintrusive ROMs in \figref{errorTime_2DAdvec}. The variation of relative error in time is similar for intrusive and nonintrusive ROMs for $r = 20$ modes. For $r = 30$ modes, the errors for nonintrusive ROMs are higher than those for intrusive ROMs. These observations are also echoed in the plot of variation in total error against the number of modes as shown in \figref{Totalerror_2Dadvec}. The results indicate that intrusive and nonintrusive ROMs exhibit similar errors for a lower number of modes. With an increase in the number of modes, intrusive ROMs exhibit a monotonous decrease in the error, while errors for nonintrusive ROMs asymptote to $10^{-3}$ for a higher number of modes. As discussed for the other two numerical experiments, this behavior indicates the reduced consistency in nonintrusive ROMs compared to FOMs. These results highlight the success of S-LDOs in enabling nonintrusive model reduction for 2-D hyperbolic PDEs.

\section{Conclusions}
\label{sec:Conclusions}

Projection-based ROMs are integral in long-term dynamics prediction, real-time control, and multi-query applications, where FOMs become very expensive. However, these ROMs are typically nonintrusive, which prohibits their use in situations where access to the source code of the underlying FOM is not available. Such situations are common when commercial simulation software packages or confidential software programs for national security applications are under consideration. For these scenarios, accurate and stable nonintrusive projection-based ROMs are immensely important. 

This article investigated using S-LDOs to enable nonintrusive projection-based ROMs. S-LDOs are ideally suited for nonintrusive ROMs as they can be directly integrated into existing intrusive ROM frameworks. Furthermore, S-LDOs enable nonintrusive model reduction even for Petrov-Galerkin projection-based ROMs. We also prove the theoretical stability of Galerkin projection-based ROMs in conjunction with S-LDOs. The applicability of S-LDO enabled nonintrusive Galerkin and LSPG projection-based ROMs by evaluating it for three numerical experiments: 1-D advection, 1-D Burgers and 2-D advection. We observed that nonintrusive Galerkin projection-based ROMs obtained using S-LDO yield ROM operators similar to the intrusive ones for fewer reduced states. Consequently, results obtained using nonintrusive ROMs yield similar accuracy as intrusive ROMs for all these problems for these lower number of modes. At a higher number of modes, errors for nonintrusive ROMs are higher than for intrusive ROMs. This observation relates to a reduction in consistency while using nonintrusive ROMs, reinstating the tradeoff between the flexibility of nonintrusive ROMs and the consistency of intrusive ROMs. 

While we demonstrate S-LDOs to be a powerful approach for enabling nonintrusive projection-based ROMs, additional work is needed to demonstrate the applicability and scalability of the approach for other problems. We plan to integrate S-LDOs with existing scalable ROM frameworks to enable nonintrusive ROM capabilities in those frameworks. This effort will also allow validation of the applicability of S-LDOs for nonintrusive model reduction for different 3D complex problems that may use unstructured grids. For higher solution accuracy in complex applications involving boundary layers or shock waves, S-LDOs must be improved. For such scenarios, the stability constraints may yield a suboptimal differential operator that does not capture these physical features. Consequently, another research direction for the future is to identify alternate stability constraints for S-LDOs that would yield a more accurate solution representation.

\section{Acknowledgements}

The authors would like to acknowledge the support from the National Science Foundation (NSF) grants, CMMI-1953323 and CBET-2332084, for the funds used towards this project. The research in this paper was also sponsored by the Army Research Laboratory and was accomplished under Cooperative Agreement Number W911NF-20-2-0175. The views and conclusions contained in this document are those of the authors and should not be interpreted as representing the official policies, either expressed or implied, of the Army Research Laboratory or the U.S. Government. The U.S. Government is authorized to reproduce and distribute reprints for Government purposes notwithstanding any copyright notation herein.

\bibliographystyle{unsrt}
\bibliography{main_bbl.bib}

\begin{thebibliography}{10}

\bibitem{Lumley1967}
J.~L. Lumley.
\newblock The structure of inhomogeneous turbulent flows.
\newblock In {\em Atmospheric Turbulence and Radio Wave Propagation}, pages 166--178. Moscow: Nauka, 1967.

\bibitem{Hesthaven2016}
J.~S. Hesthaven, G.~Rozza, and B.~Stamm.
\newblock {\em Certified Reduced Basis Methods for Parametrized Partial Differential Equations}.
\newblock Springer, 2016.

\bibitem{Lee2020}
K.~Lee and K.~T. Carlberg.
\newblock Model reduction of dynamical systems on nonlinear manifolds using deep convolutional autoencoders.
\newblock {\em Journal of Computational Physics}, 404:108973, 2020.

\bibitem{Kim2022}
Y.~Kim, Y.~Choi, D.~Widemann, and T.~Zohdi.
\newblock A fast and accurate physics-informed neural network reduced order model with shallow masked autoencoder.
\newblock {\em Journal of Computational Physics}, 451:110841, 2022.

\bibitem{Puri2024}
V.~Puri, A.~Prakash, L.~B. Kara, and Y.~J. Zhang.
\newblock {SNF-ROM: Projection-based nonlinear reduced order modeling with smooth neural fields}.
\newblock {\em arXiv: 2405.14890}, 2024.

\bibitem{Aubry1988}
N.~Aubry, P.~Holmes, J.~L. Lumley, and E.~Stone.
\newblock The dynamics of coherent structures in the wall region of a turbulent boundary layer.
\newblock {\em Journal of Fluid Mechanics}, 192:115–173, 1988.

\bibitem{Carlberg2011}
K.~Carlberg, C.~Bou-Mosleh, and C.~Farhat.
\newblock {Efficient non-linear model reduction via a least-squares Petrov–Galerkin projection and compressive tensor approximations}.
\newblock {\em International Journal for Numerical Methods in Engineering}, 86(2):155--181, 2011.

\bibitem{Carlberg2017}
K.~Carlberg, M.~Barone, and H.~Antil.
\newblock {Galerkin v. least-squares Petrov–Galerkin projection in nonlinear model reduction}.
\newblock {\em Journal of Computational Physics}, 330:693--734, 2017.

\bibitem{Parish2020}
E.~J. Parish, C.~R. Wentland, and K.~Duraisamy.
\newblock {The adjoint Petrov–Galerkin method for non-linear model reduction}.
\newblock {\em Computer Methods in Applied Mechanics and Engineering}, 365:112991, 2020.

\bibitem{Parish2023}
E.~J. Parish, M.~Yano, I.~Tezaur, and T.~Iliescu.
\newblock Residual-based stabilized reduced-order models of the transient convection-diffusion-reaction equation obtained through discrete and continuous projection.
\newblock {\em arXiv: 2302.09355}, 2023.

\bibitem{Peherstorfer2016}
B.~Peherstorfer and K.~E. Willcox.
\newblock Data-driven operator inference for nonintrusive projection-based model reduction.
\newblock {\em Computer Methods in Applied Mechanics and Engineering}, 306:196--215, 2016.

\bibitem{Hesthaven2018}
J.~S. Hesthaven and S.~Ubbiali.
\newblock Non-intrusive reduced order modeling of nonlinear problems using neural networks.
\newblock {\em Journal of Computational Physics}, 363:55--78, 2018.

\bibitem{Fries2022}
W.~D. Fries, X.~He, and Y.~Choi.
\newblock Lasdi: Parametric latent space dynamics identification.
\newblock {\em Computer Methods in Applied Mechanics and Engineering}, 399:115436, 2022.

\bibitem{Padovan2024}
A.~Padovan, B.~Vollmer, and D.~J. Bodony.
\newblock Data-driven model reduction via non-intrusive optimization of projection operators and reduced-order dynamics.
\newblock {\em arXiv: 2401.01290}, 2024.

\bibitem{Rowley2009}
C.~W. Rowley, I.~Mezic, S.~Bagheri, P.~Schlatter, and D.~S. Henningson.
\newblock Spectral analysis of nonlinear flows.
\newblock {\em Journal of Fluid Mechanics}, 641:115–127, 2009.

\bibitem{Schmid2010}
P.~J. Schmid.
\newblock Dynamic mode decomposition of numerical and experimental data.
\newblock {\em Journal of Fluid Mechanics}, 656:5–28, 2010.

\bibitem{Schmid2022}
P.~J. Schmid.
\newblock Dynamic mode decomposition and its variants.
\newblock {\em Annual Review of Fluid Mechanics}, 54:225--254, 2022.

\bibitem{Champion2019}
K.~Champion, B.~Lusch, J.~N. Kutz, and S.~L. Brunton.
\newblock Data-driven discovery of coordinates and governing equations.
\newblock {\em Proceedings of the National Academy of Sciences}, 116(45):22445--22451, 2019.

\bibitem{Kalashnikova2010}
I.~Kalashnikova and M.~F. Barone.
\newblock {On the stability and convergence of a Galerkin reduced order model (ROM) of compressible flow with solid wall and far-field boundary treatment}.
\newblock {\em International Journal for Numerical Methods in Engineering}, 83(10):1345--1375, 2010.

\bibitem{Ingimarson2022}
S.~Ingimarson, L.~G. Rebholz, and T.~Iliescu.
\newblock Full and reduced order model consistency of the nonlinearity discretization in incompressible flows.
\newblock {\em Computer Methods in Applied Mechanics and Engineering}, 401:115620, 2022.

\bibitem{McQuarrie2021}
S.~A. McQuarrie, C.~Huang, and Karen~E. Willcox.
\newblock Data-driven reduced-order models via regularised operator inference for a single-injector combustion process.
\newblock {\em Journal of the Royal Society of New Zealand}, 51(2):194--211, 2021.

\bibitem{Kramer2024}
B.~Kramer, B.~Peherstorfer, and K.~E. Willcox.
\newblock Learning nonlinear reduced models from data with operator inference.
\newblock {\em Annual Review of Fluid Mechanics}, 56(1):null, 2024.

\bibitem{Peherstorfer2020}
B.~Peherstorfer.
\newblock Sampling low-dimensional markovian dynamics for preasymptotically recovering reduced models from data with operator inference.
\newblock {\em SIAM Journal on Scientific Computing}, 42(5):A3489--A3515, 2020.

\bibitem{Sawant2023}
N.~Sawant, B.~Kramer, and B.~Peherstorfer.
\newblock Physics-informed regularization and structure preservation for learning stable reduced models from data with operator inference.
\newblock {\em Computer Methods in Applied Mechanics and Engineering}, 404:115836, 2023.

\bibitem{Sharma2022}
H.~Sharma, Z.~Wang, and B.~Kramer.
\newblock {Hamiltonian operator inference: Physics-preserving learning of reduced-order models for canonical Hamiltonian systems}.
\newblock {\em Physica D: Nonlinear Phenomena}, 431:133122, 2022.

\bibitem{Grimberg2020}
S.~Grimberg, C.~Farhat, and N.~Youkilis.
\newblock On the stability of projection-based model order reduction for convection-dominated laminar and turbulent flows.
\newblock {\em Journal of Computational Physics}, 419:109681, 2020.

\bibitem{Schumann2022}
Y.~Schumann and P.~Neumann.
\newblock Towards data-driven inference of stencils for discrete differential operators.
\newblock In {\em Proceedings of the Platform for Advanced Scientific Computing Conference}, New York, NY, USA, 2022.

\bibitem{Schumann2023}
Y.~Schumann and P.~Neumann.
\newblock On linear models for discrete operator inference in time dependent problems.
\newblock {\em Journal of Computational and Applied Mathematics}, 425:115022, 2023.

\bibitem{Gkimisis2023}
L.~Gkimisis, T.~Richter, and P.~Benner.
\newblock Adjacency-based, non-intrusive reduced-order modeling for fluid-structure interactions.
\newblock {\em Proceedings in Applied Mathematics and Mechanics}, 23(4), 2023.

\bibitem{Gkimisis2024}
L.~Gkimisis, T.~Richter, and P.~Benner.
\newblock Adjacency-based, non-intrusive model reduction for vortex-induced vibrations.
\newblock {\em Computers \& Fluids}, 275:106248, 2024.

\bibitem{Prakash2024b}
A.~Prakash and Y.~J. Zhang.
\newblock Data-driven identification of stable sparse differential operators using constrained regression.
\newblock {\em Computer Methods in Applied Mechanics and Engineering}, 429:117149, 2024.

\bibitem{LeVeque2007}
R.~J. LeVeque.
\newblock {\em Finite Difference Methods for Ordinary and Partial Differential Equations}.
\newblock Society for Industrial and Applied Mathematics, 2007.

\bibitem{Duff2002}
I.~S. Duff, M.~A. Heroux, and R.~Pozo.
\newblock An overview of the sparse basic linear algebra subprograms: The new standard from the blas technical forum.
\newblock {\em ACM Trans. Math. Softw.}, 28(2):239–267, jun 2002.

\bibitem{Charney1950}
J.~G. Charney, R.~Fjortoft, and J.~V.~Neumann.
\newblock Numerical integration of the barotropic vorticity equation.
\newblock {\em Tellus}, 2(4):237--254, 1950.

\bibitem{Isaacson1994}
E.~Isaacson and H.~B. Keller.
\newblock {\em Analysis of Numerical Methods}.
\newblock Dover Publications, Inc., 1994.

\bibitem{Evans2023}
J.~A. Evans.
\newblock Chapter 4 - spline-based methods for turbulence.
\newblock In Robert~D. Moser, editor, {\em Numerical Methods in Turbulence Simulation}, Numerical Methods in Turbulence, pages 139--187. Academic Press, 2023.

\bibitem{LeVeque2002}
R.~J. LeVeque.
\newblock {\em Finite Volume Methods for Hyperbolic Problems}.
\newblock Cambridge Texts in Applied Mathematics. Cambridge University Press, 2002.

\bibitem{Sastry1999}
S.~Sastry.
\newblock {\em Lyapunov Stability Theory}, pages 182--234.
\newblock Springer New York, 1999.

\bibitem{Rouche1977}
N.~Rouche, P.~Habets, and M.~Laloy.
\newblock {\em Stability Theory by Liapunov’s Direct Method}.
\newblock Springer New York, NY, 1977.

\bibitem{Boots2007}
B.~Boots, G.~J. Gordon, and S.~Siddiqi.
\newblock A constraint generation approach to learning stable linear dynamical systems.
\newblock In {\em Advances in Neural Information Processing Systems}, volume~20. Curran Associates, Inc., 2007.

\bibitem{Goyal2023}
P.~Goyal, I.~P. Duff, and P.~Benner.
\newblock Inference of continuous linear systems from data with guaranteed stability.
\newblock {\em arXiv: 2301.10060}, 2023.

\bibitem{Goyal2024}
P.~Goyal, I.~P. Duff, and P.~Benner.
\newblock {Guaranteed stable quadratic models and their applications in SINDy and operator inference}.
\newblock {\em arXiv: 2308.13819}, 2023.

\bibitem{Kalashnikova2014}
I.~Kalashnikova, M.~F. Barone, S.~Arunajatesan, and B.~G. V.~B. Waanders.
\newblock Construction of energy-stable projection-based reduced order models.
\newblock {\em Applied Mathematics and Computation}, 249:569--596, 2014.

\bibitem{Chen2023}
P.~Y. Chen, J.~Xiang, D.~H. Cho, Y.~Chang, G.~A. Pershing, H.~T. Maia, M.~M. Chiaramonte, K.~T. Carlberg, and E.~Grinspun.
\newblock {CROM}: Continuous reduced-order modeling of {PDE}s using implicit neural representations.
\newblock In {\em The Eleventh International Conference on Learning Representations}, 2023.

\bibitem{Barnett2022}
J.~Barnett and C.~Farhat.
\newblock {Quadratic approximation manifold for mitigating the Kolmogorov barrier in nonlinear projection-based model order reduction}.
\newblock {\em Journal of Computational Physics}, 464:111348, 2022.

\bibitem{Geelen2023}
R.~Geelen, S.~Wright, and K.~Willcox.
\newblock Operator inference for non-intrusive model reduction with quadratic manifolds.
\newblock {\em Computer Methods in Applied Mechanics and Engineering}, 403:115717, 2023.

\bibitem{Hughes2000}
T.~J.~R. Hughes.
\newblock {\em The Finite Element Method: Linear Static and Dynamic Finite Element Analysis}.
\newblock Dover Publications, Inc., 2000.

\bibitem{Iliescu2014}
T.~Iliescu and Z.~Wang.
\newblock {Variational multiscale proper orthogonal decomposition: Navier-Stokes equations}.
\newblock {\em Numerical Methods for Partial Differential Equations}, 30(2):641--663, 2014.

\bibitem{Ahmed2021}
S.~E. Ahmed, S.~Pawar, O.~San, A.~Rasheed, T.~Iliescu, and B.~R. Noack.
\newblock {On closures for reduced order models - A spectrum of first-principle to machine-learned avenues}.
\newblock {\em Physics of Fluids}, 33(9), 2021.

\bibitem{Prakash2024a}
A.~Prakash and Y.~J. Zhang.
\newblock Projection-based reduced order modeling and data-driven artificial viscosity closures for incompressible fluid flows.
\newblock {\em Computer Methods in Applied Mechanics and Engineering}, 425:116930, 2024.

\bibitem{Rowley2004}
C.~W. Rowley, T.~Colonius, and R.~M. Murray.
\newblock {Model reduction for compressible flows using POD and Galerkin projection}.
\newblock {\em Physica D: Nonlinear Phenomena}, 189(1):115--129, 2004.

\bibitem{Barone2009}
M.~F. Barone, I.~Kalashnikova, D.~J. Segalman, and H.~K. Thornquist.
\newblock {Stable Galerkin reduced order models for linearized compressible flow}.
\newblock {\em Journal of Computational Physics}, 228(6):1932--1946, 2009.

\bibitem{Kalashnikova2014b}
I.~Kalashnikova, B.~{V. B. Waanders}, S.~Arunajatesan, and M.~Barone.
\newblock Stabilization of projection-based reduced order models for linear time-invariant systems via optimization-based eigenvalue reassignment.
\newblock {\em Computer Methods in Applied Mechanics and Engineering}, 272:251--270, 2014.

\bibitem{Amsallem2012}
D.~Amsallem and C.~Farhat.
\newblock Stabilization of projection-based reduced-order models.
\newblock {\em International Journal for Numerical Methods in Engineering}, 91(4):358--377, 2012.

\bibitem{Qian2022}
E.~Qian, I.~Farca\c{s}, and K.~Willcox.
\newblock Reduced operator inference for nonlinear partial differential equations.
\newblock {\em SIAM Journal on Scientific Computing}, 44(4):A1934--A1959, 2022.

\bibitem{Rezaian2021}
E.~Rezaian and M.~Wei.
\newblock A global eigenvalue reassignment method for the stabilization of nonlinear reduced-order models.
\newblock {\em International Journal for Numerical Methods in Engineering}, 122(10):2393--2416, 2021.

\bibitem{Tomoki2024}
T.~Koike and E.~Qian.
\newblock Energy-preserving reduced operator inference for efficient design and control.
\newblock In {\em AIAA SCITECH Forum}, number AIAA 2024-1012, 2024.

\bibitem{2020SciPy}
{P. Virtanen, R. Gommers, T. E. Oliphant, M. Haberland, T. Reddy \textit{et al.}}
\newblock {SciPy} 1.0: Fundamental algorithms for scientific computing in {P}ython.
\newblock {\em Nature Methods}, 17:261--272, 2020.

\end{thebibliography}

\end{document}